\documentclass[12pt,a4paper]{article}
\usepackage{amsmath}  % an extension package for LaTeX that provides various features to facilitate writing math formulas and to improve the typographical quality of their output
\usepackage{amssymb}  % 
\usepackage{graphicx} % importing images into the document 
\usepackage{bm}       % allows to write bold greek letters
\usepackage{geometry} % to change page margins 
\usepackage{multirow} % to create multirow table headers
\usepackage{times}    % to use Times font
\usepackage[dvips,colorlinks,bookmarksopen,bookmarksnumbered,citecolor=red,urlcolor=red]{hyperref}
\title{Adaptive Markov Chain Monte Carlo Forward Simulation for Statistical Analysis in Epidemic Modelling of Human Papillomavirus.}

\author{Igor A.~Korostil\thanks{The {K}irby {I}nstitute, University of {N}ew {S}outh {W}ales, Australia}\and Gareth W.~Peters\thanks{Mathematics and Statistics
Department, University of {N}ew {S}outh {W}ales, Australia.} \and Julien Cornebise\thanks{Department of Statistics, UCL, London, UK.; J.C. is supported by {BBSRC} grant BB/G006997/1}\and David G.~Regan\thanks{The {K}irby {I}nstitute, University of {N}ew {S}outh {W}ales, Australia}}
\begin{document}

\maketitle

\begin{abstract}
\noindent  We develop a Bayesian statistical model and estimation methodology based on Forward Projection Adaptive Markov chain Monte Carlo in order to perform the calibration of a high-dimensional non-linear system of Ordinary Differential Equations representing an epidemic model for Human Papillomavirus types 6 and 11 (HPV-6, HPV-11). The model is compartmental and involves stratification by age, gender and sexual activity-group. Developing this model and a means to calibrate it efficiently is relevant since HPV is a very multi-typed and common sexually transmitted infection with more than 100 types currently known. The two types studied in this paper, types 6 and 11, are causing about 90\% of anogenital warts.

We extend the development of a sexual mixing matrix for the population, based on a formulation first suggested by Garnett and Anderson. In particular we consider a stochastic mixing matrix framework which allows us to jointly estimate unknown attributes and parameters of the mixing matrix along with the parameters involved in the calibration of the HPV epidemic model. This matrix describes the sexual interactions between members of the population under study and relies on several quantities which are \textit{a priori} unknown. The Bayesian model developed allows one to estimate jointly the HPV-6 and HPV-11 epidemic model parameters such as the probability of transmission, HPV incubation period, duration of infection, duration of genital warts treatment, duration of immunity, the probability of seroconversion, per gender, age-group and sexual activity-group, as well as unknown sexual mixing matrix parameters related to assortativity.

Finally, we explore the ability of an extension to the class of adaptive Markov chain Monte Carlo algorithms to incorporate a forward projection simulation strategy for the ordinary differential equation state trajectories. Efficient exploration of the Bayesian posterior distribution developed for the ODE parameters provides a challenge for any Markov chain sampling methodology, hence the interest in adaptive Markov chain methods. We conclude with simulation studies on synthetic and actual data from studies undertaken recently in Australia.
\end{abstract}

\section{Background on Modelling HPV and Relevance to Community Health.}
The human papillomaviruses (HPV) are a family of small DNA viruses that preferentially infect differentiating epithelial cells of the skin and mucosae. More than 100 HPV genotypes have thus far been identified, classified according to their tissue tropism (mucosal or cutaneous) and oncogenic potential (high or low). About 40 HPV types are known to infect the mucosae, including those of the anogenital and oral tracts, and 13-18 of these are considered to be oncogenic (high-risk) on the basis of their association with malignancies. Low-risk HPV types are associated with benign lesions such as genital warts and low-grade intraepithelial neoplasias of the cervix \cite{Munoz2006,Trottier2009}. Sexual contact is the primary mode of transmission \cite{Burchell2006a} and HPV is the most common sexually transmitted infection in the world. HPV is known to be the causal factor in the vast majority of cervical cancer cases and is also implicated in a proportion of other anogenital cancers and cancers of the head and neck. The overall burden of disease attributed to HPV, both cancers (as much as 5.2\% of incident cancers worldwide) and benign lesions such as genital warts, is considerable \cite{Lacey2006}.

Two vaccines have been developed and shown through clinical trials to be highly effective in the prevention of precancerous lesions and persistent infection due to an important subset of HPV types \cite{Pomfret2011,Stanley2010}. The quadrivalent vaccine (Gardasil) protects against high-risk HPV types 16 and 18 that are associated with 70-75\% of cervical cancers, and against low-risk HPV types 6 and 11 that cause more than 90\% of genital warts. The bivalent vaccine (Cervarix) provides protection against HPV types 16 and 18 only. Both vaccines have been licensed in more than 100 countries and publicly funded national immunisation programmes have commenced in some of these including Australia (Gardasil) \cite{Garland2010}.

National immunisation programmes are costly and decisions regarding their implementation are generally made on the basis of health-economic evaluations. In regard to HPV, these decisions are complicated by two related factors: 1) HPV is a sexually transmitted infection; and 2) only a small proportion of infections do not resolve and can lead to cancer many years or decades subsequent to acquisition. Both of these factors have generally been addressed by employing models to estimate the long-term impact of vaccination on the incidence of HPV-related disease so that the costs and benefits can be calculated. However, it is the former of these factors that is of particular relevance in the context of this study. Because HPV is an infectious disease, the rate of transmission in a population, commonly referred to as the "force of infection", is a function of the prevalence of the infection in the population at any given time \cite{Anderson1992}. Furthermore, the benefit of vaccination that confers immunity to infection (immunisation) is not confined only to those directly immunised--- "unvaccinated individuals" (who remain susceptible to infection) enjoy a degree of indirect protection because their risk of exposure is reduced through a diminishment in circulating virus. This indirect benefit of vaccination is referred to as "herd immunity" \cite{Garnett2005}. In order to model the impact of vaccination on the course of the HPV epidemic over time, in a manner that captures the herd immunity effect, we must use dynamic transmission models \cite{Brisson2003,Edmunds1999}. Failure to do so can result in an underestimation of the potential benefit of vaccination.

Dynamic mathematical transmission models have been used extensively to estimate the potential impact of HPV vaccination in a wide variety of settings (e.g., \cite{Barnabas2006,Choi2010,Elbasha2008,Hughes2002,Regan2007}, not comprehensive) and as a component of cost-effectiveness evaluations that have informed decisions on the funding of vaccination programmes (e.g., \cite{Beutels2010,Jit2008,Kim2008,Kulasingam2007,Marra2009,Sanders2003}, not comprehensive). Transmission models have traditionally been formulated in a deterministic framework as systems of differential equations (ordinary or partial) \cite{Anderson1992,Keeling2007,Vynnycky2010}. With the increasing power of personal desktop computers and access to high performance computing facilities, the use of agent-based stochastic modelling approaches has become more prominent (examples for HPV include \cite{Burchell2006,VandeVelde2007}). The latter approach is particularly useful for low prevalence infections where there is a possibility of extinction and/or where it is necessary to capture events that occur at the level of the individual (e.g., tracing and treating sexual partners of infected individuals). However, for their computational efficiency, analytical tractability and ability to provide mechanistic insights to epidemic dynamics, deterministic ordinary differential equation (ODE) models are often preferred, particularly for endemic infections such as HPV \cite{Regan2008}. We have previously developed a deterministic single-type transmission model for HPV-16 \cite{Regan2007} and more recently multi-type models for HPV types 6, 11, 16 and 18 in order to evaluate the potential impact of vaccination. In this paper we discuss a novel formulation of a model for HPV-6 and -11 that incorporates Australian data on genital warts incidence  and type-specific seroprevalence.

In this study, we develop a Bayesian statistical model and estimation methodology to perform the calibration of a high-dimensional non-linear system of ordinary differential equations representing an epidemic model for HPV types 6 and 11. While the health and economic consequences of HPV-16 and -18 are more serious than for types 6 and 11 (hence their inclusion in both currently available vaccines), a model for types 6 and 11 was chosen for this study because of the availability of Australian genital warts incidence \cite{Pirotta2009} and seroprevalence \cite{Newall2008} data that could be used for calibration.

Despite extensive study, there remains considerable uncertainty regarding aspects of the natural history of HPV and the patterns of sexual behaviour that underpin transmission \cite{Regan2010,Veldhuijzen2010}. Furthermore, many studies have not been designed with transmission models in mind, so the processes and phenomena they measure cannot always be applied directly and/or their interpretation in the context of transmission is not clear. Areas of uncertainty that present particular challenges for modelling include interpretation of vaccine efficacy from clinical trials in the context of transmission, the duration of infectiousness (as opposed to the duration of detectability using currently available tests), the duration and nature of naturally acquired immunity, the relationship between seropositivity and immunity, and the probability of transmission on sexual contact. Some of these are difficult to measure at a population level for practical and/or ethical reasons. We demonstrate a Bayesian statistical methodology that will address these uncertainties in the estimation and calibration of the ODE epidemic model we have developed. This methodology allows us to statistically quantify the extent of the uncertainty in model outcomes that are derived from uncertainty in the inputs, and the contribution of uncertainty in individual parameters to the uncertainty in the outcomes \cite{Saltelli2000, Hoare2008}. It can also be used to predict the impact of vaccination on a population.  

\subsection{Introduction to Adaptive Markov chain Monte Carlo}
Markov chain Monte Carlo (MCMC) sampling has gained a wide recognition in all areas of modelling and statistical estimation as an essential tool for performing inference in Bayesian models (see reviews and discussions in \cite{gilks1996markov} and \cite{brooks1998markov}). In this paper we consider the recently developed class of algorithms known as Adaptive MCMC (see a review in \cite{Andrieu2008}), and demonstrate how they may be extended to solve statistically challenging estimation and prediction problems of direct relevance to the interpretation and analysis of the calibration and vaccine response dynamics for HPV epidemic models. We illustrate this on the model we develop for HPV-6 and -11.

Standard MCMC algorithms that do not incorporate adaptation often require a degree of "tuning" of the parameters controlling the algorithms performance. This is typically performed by off-line simulations to assess performance of the mixing of the resulting Markov chain followed by numerical investigation of the convergence rates to stationarity of the chain for different algorithmic settings of the proposal distribution. For example, the widely used variant of the Metropolis Hastings algorithm, the Random Walk Metropolis algorithm has mixing performance that is controlled through specification of the Markov chain proposal distributions covariance matrix. Tuning this matrix for optimal performance can be computationally expensive and inefficient. Optimal performance of an MCMC algorithm is typically either specified by the convergence rate of the Markov chain to stationarity or through the related quantity, the acceptance probability of the rejection step in the MCMC algorithm. In this regard, theoretically optimal results have been derived for several classes of statistical models, which now act as guides for more complicated sampling problems (see discussions in \cite{roberts2001optimal}).

In this paper, the ODE HPV epidemic model is constructed on a high dimensional space both in the parameters of the model and also in the latent ODE state trajectories solved for at each discrete time point in the "forward projection" ODE solver. This high dimensionality in the posterior parameter space provides a significant challenge for standard MCMC algorithms with respect to the design of an efficient proposal mechanism for the Markov chain. In particular, in the model considered in this paper the fact that we incorporate a forward projection stage for the ODE solver adds additional complications in the design of the proposal. Therefore, it is desirable to automate this proposal construction for the MCMC sampler, avoiding computationally expensive tuning processes. Hence, we develop an adaptive version of the Random Walk Metropolis algorithm, coupled with forward projection. The incorporation of an adaptive proposal mechanism in an Markov chain Monte Carlo algorithm has been demonstrated to improve the performance of the sampling algorithm relative to standard MCMC approaches, see reviews of several examples of this improvement in \ref{Andrieu2008}. This improvement is achieved by learning on-line the structure of the Markov chain proposal distribution in an automated fashion, avoiding tuning of the MCMC proposal mechanism.

There are several classes of adaptive MCMC algorithms and each class has several adaptation strategies \cite{Atchade2005, Andrieu2006}. These approaches can be classified as either internal adaptation mechanisms, including controlled MCMC methods or external adaptation strategies (see discussion in \cite{Atchade2005}). The distinguishing feature of adaptive MCMC algorithms, when compared to standard MCMC, is that the Markov chain is generated via a sequence of transition kernels. Adaptive algorithms get their name from the fact that they utilise a combination of time or state inhomogeneous proposal kernels. Each proposal in the sequence is allowed to depend on the past history of the Markov chain generated, resulting in many possible variants. When using inhomogeneous Markov kernels it is particularly important to ensure the generated Markov chain is ergodic, with the appropriate stationary distribution. Several recent papers proposing theoretical conditions that must be satisfied to ensure ergodicity of adaptive algorithms include \cite{Atchade2005} and \cite{Haario2001}. The paper  \cite{Roberts2009} proves ergodicity of adaptive MCMC under conditions known as \textit{Diminishing Adaptation} and \textit{Bounded Convergence}. Designing an adaptation strategy that satisfies these conditions guarantees asymptotic convergence of the law of the Markov chain samples to the target posterior and ensures the Weak Law of Large Numbers holds for bounded test functions of the parameter space (an interested reader is referred to \cite{Roberts2009} for details). In this paper we work with a transition kernel that is well known to satisfy these conditions and has been used successfully in several applications, based on the adaptive Metropolis algorithm.

%\vspace{-2cm}
\subsection{Contributions}
In this paper we formulated a model for the HPV types 6 and 11 which adequately covers all aspects of the disease transmission and treats the incorporated seropositivity as a state associated with an individual's recovery. Our aim was to ensure the model structure is minimalistic yet sufficient for testing whether this particular interpretation of seropositivity agrees with the available data the model is calibrated to.

Using this deterministic ODE model we undertake three key tasks: construction of a robust statistical modelling framework under a Bayesian paradigm to perform calibration of this coupled ODE transmission model with extensions to a stochastic population mixing matrix formulation; development of an automated statistical estimation methodology, based on a modification to adaptive MCMC to incorporate forward projection for the ODE, that will provide a robust means of performing calibration and statistical analysis of the calibration performance; and a statistical methodology to study vaccination responses based on the posterior predictive distribution we estimated via adaptive MCMC.

We perform detailed studies on synthetic data to assess the properties of the model and adaptive MCMC forward projection methodology. This is followed by assessment of the calibration performance on real data collected from Australian sources (genital warts incidence \cite{Pirotta2009}, and HPV-6 and -11 type-specific seroprevalence \cite{Newall2008}).

\section{Human Papillomavirus (HPV) Transmission Model}
\label{HPVModel}

\begin{figure}[h!]
\begin{center}
\includegraphics[scale = 1.2]{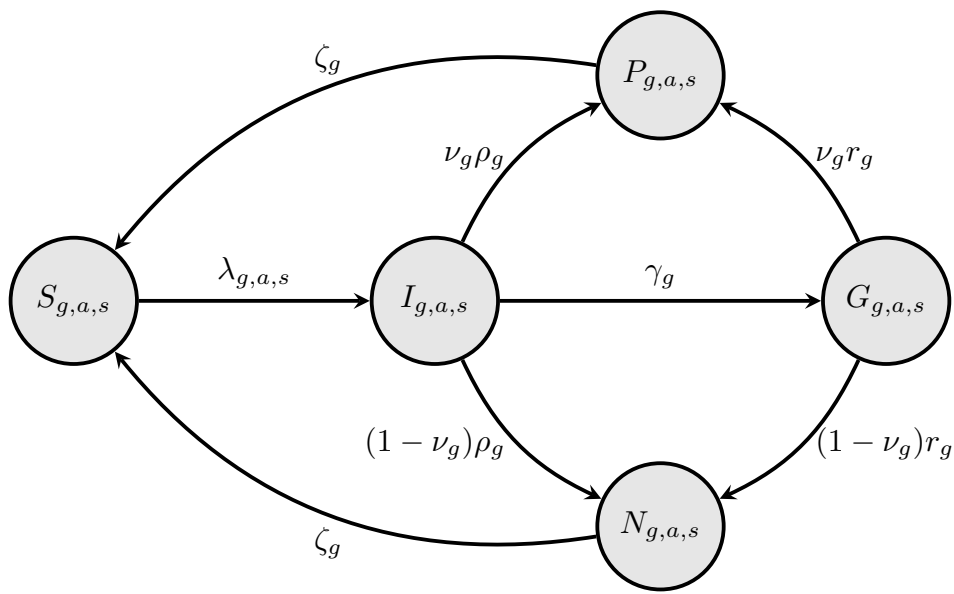}
\caption{A compartmental HPV-6/-11 transmission model (see Table \ref{tab:compartments}).}
\label{fig:genital_warts_model}
\end{center}
\end{figure}

We consider a dynamic transmission model for HPV types 6 and 11 (Figure \ref{fig:genital_warts_model}). This model is compartmental, such that the entire population is viewed as being distributed between a set of non-overlapping groups ("compartments") representing the stages of disease progression. The compartments are specified as presented in Table \ref{tab:compartments}.  The model is intended to describe how the number of people in each compartment changes over time. For example, members of the susceptible population 'move' from $S$ to $I$ as they become infected, and members of the recovered seropositive population 'move' from $P$ to $S$ as they lose their immunity.
\subsection*{Underlying Epidemic Model Assumptions}
Here we briefly discuss the key assumptions the model is based upon.
%%%%%%%%%%%%%%%%%%%%%%%%%%%%%%%%%
%%%%%%%%%%%%%%%%%%%%%%%%%%%%%%%%%
\begin{description}
%%%%%%%%%%%%%%%%%%%%%%%%%%%%%%%%%
%%%%%%%%%%%%%%%%%%%%%%%%%%%%%%%%%
 	\item[Population] (a) the modelled population consists of people aged 15-59 y.o. who are divided into 9 separate five-year age-groups (see Table \ref{tab:relative_rate_by_age_group}).
Limiting the population to this particular age range is motivated by a presumed absence of sexual activity in people younger than 15 y.o. and people over 59, though any extensions to these ranges is not precluded by our methodology; five-year age-groups are  commonly used for reporting results of surveys and trials including the one providing the sexual behaviour data for our model (see, for example, \cite{Pirotta2009,Newall2008,Giuliano2008});  (b) the modelled population is constant over time; consequently, immigrants and temporary visitors are not accounted for, which may be an important simplification for Australia whose population has been steadily growing due to immigration; furthermore, Australia is a popular destination for young travellers who often maintain a high level of sexual activity; (c) mortality, though formally implemented, is not caused by HPV infection but rather serves as a convenient way of removing individuals who do not contribute to the HPV transmission due to advanced age signified by a cessation of sexual activity; (d) the number of males in the whole population and every sexual activity or age-group is equal to the number of females; (e)  no transition between genders is allowed (i.e. males can not become females and vice versa);
%%%%%%%%%%%%%%%%%%%%%%%%%%%%%%%%%
%%%%%%%%%%%%%%%%%%%%%%%%%%%%%%%%%
\item[Sexual behaviour] (a) the modelled population is heterosexual with all people belonging to one of four sexual activity (risk) groups (group one being the least active and group four the most active); the proportions of the population in each risk-group 1-4 are 0.6, 0.27, 0.11 and 0.02, respectively \cite{Smith2003,Regan2007} (also see Table \ref{tab:relative_rate_by_sex_group}); (b) people are born into a particular risk-group and can never leave this group but their activity level is a function of their age; this restriction implies, for example, that even when someone from the most active group gets older, his or her activity level declines but does not drop to the level of representatives of a less active group of the same age; (c) the annual sexual partner change rate for an individual is fully defined by his or her age and risk-group; the manner in which an individual of a given age, gender and risk-group "chooses" partners of the opposite gender and of a given age and risk-group is described by means of a sexual mixing matrix (discussed in section \ref{sec:smm});
%%%%%%%%%%%%%%%%%%%%%%%%%%%%%%%%%
%%%%%%%%%%%%%%%%%%%%%%%%%%%%%%%%%
\item[HPV transmission and seropositivity] (a) we assume people seek treatment immediately upon becoming aware that they have genital warts; (b) we associate seropositivity and seronegativity exclusively with the recovered state such that only those in the recovered (immune) state can have the status of seropositive/seronegative.This status is lost upon removal from the recovered state to the susceptible state (i.e., loss of immunity). In the context of our model the recovered/seropositive state corresponds to those people who have developed detectable antibodies (using currently available tests) to to HPV-6/-11 through an immunogenic response. Conversely, the recovered/seronegative state corresponds to those who have recovered and are immune to reinfection but have not developed detectable antibodies.  In general, seropositivity can serve as a long-lasting marker of ongoing or prior infection, though not a particularly reliable one in the case of HPV as only a proportion of those exposed to infection develop detectable antibodies \cite{Newall2008}. Furthermore, there is some evidence supporting an alternative notion of seropositiviy to the one we have assumed, whereby seropositivity in only a marker of previous infection and is not correlated with protection against reinfection (see \cite{Trottier2008}). Such a perspective would lead to a more complicated model structure and we therefore do not focus on it in this paper. However, the methodology we develop can easily be extended to this context.
%%%%%%%%%%%%%%%%%%%%%%%%%%%%%%%%%
%%%%%%%%%%%%%%%%%%%%%%%%%%%%%%%%%
\end{description}
%%%%%%%%%%%%%%%%%%%%%%%%%%%%%%%%%
%%%%%%%%%%%%%%%%%%%%%%%%%%%%%%%%%

\subsection*{Formulation of the model as a system of ODEs}
Our model is formulated as the following system of ordinary differential equations (ODEs):
\begin{eqnarray}
\dot{S}_{g,s,a} &=& -\lambda_{g,s,a}(t)S_{g,s,a} + \zeta_g (P_{g,s,a}+N_{g,s,a})+\frac{1}{5}S_{g,s,(a-1)}-\frac{1}{5}S_{g,s,a}+\nonumber \\
& & \frac{1}{40}\sum_{g,s}(S_{g,s,9}+I_{g,s,9}+G_{g,s,9}+P_{g,s,9}+N_{g,s,9})\delta_1(a)\times\\
&&(0.6\delta_1(s)+0.27\delta_2(s)+0.11\delta_3(s)+0.02\delta_4(s)), \label{eq:gw_system_eq1}\\
\dot{I}_{g,s,a} &=& \lambda_{g,s,a}(t)S_{g,s,a}-(\gamma_g+\rho_g)I_{g,s,a} + \frac{1}{5}I_{g,s,(a-1)}-\frac{1}{5}I_{g,s,a},\label{eq:gw_system_eq2}\\
\dot{G}_{g,s,a}&=&\gamma_g I_{g,s,a}-r_gG_{g,s,a}+\frac{1}{5}G_{g,s,(a-1)}-\frac{1}{5}G_{g,s,a},\label{eq:gw_system_eq3}\\
\dot{P}_{g,s,a}&=&\nu_g (\rho_gI_{g,s,a}+r_gG_{g,s,a})-\zeta_g P_{g,s,a}+\frac{1}{5}P_{g,s,(a-1)}-\frac{1}{5}P_{g,s,a},\label{eq:gw_system_eq4}\\
\dot{N}_{g,s,a}&=&(1-\nu_g) (\rho_gI_{g,s,a}+r_gG_{g,s,a})-\zeta_g N_{g,s,a}+\frac{1}{5}N_{g,s,(a-1)}-\frac{1}{5}N_{g,s,a}. \label{eq:gw_system_eq5}
\label{odeModel}
\end{eqnarray}
Here the capital letters denote the number of people in a compartment and the subscripts denote gender ($g$; for males $g=1$, for females $g=2$), one of the four sexual activity-groups mentioned previously ($s\in \{1,\ldots,4\}$), and an age-group ($a\in \{1,\ldots,9\}$); the dot denotes a derivative with respect to time; $\delta_1(a)$ is the Kronecker delta function, equal to 1 if $a=1$ or 0 otherwise, and the system coefficients should be interpreted as described in Table \ref{tab:parameter_interpretation}. It should be emphasised that all coefficients in the model formulation are gender specific, i.e., they can take different values for males and females.
System (\ref{eq:gw_system_eq1})-(\ref{eq:gw_system_eq5}) contains a number of terms describing the process of aging. Since each age-group comprises a five-year band, members of the population age (i.e. move to the next age-group) at a yearly rate of $1/5$. To maintain a constant population size we assume that there is an inflow of people into the susceptible compartment of the youngest age-group (group 1) as defined by:
$$
\frac{1}{40}\sum_{g,s}(S_{g,s,9}+I_{g,s,9}+G_{g,s,9}+P_{g,s,9}+N_{g,s,9})\delta_1(a)(0.6\delta_{s,1}+0.27\delta_2(s)+0.11\delta_3(s)+0.02\delta_4(s)).
$$
This term is obtained by dividing the total number of individuals leaving the oldest age-group (group 9) each year on reaching age 60, $(S_{g,s,9}+I_{g,s,9}+G_{g,s,9}+P_{g,s,9}+N_{g,s,9})/5$,  evenly between 2 genders and 4 sexual activity-groups. To every $g$ and $s$ is added $S_{g,s,1}$  according to the previously defined distribution of the population across risk-groups (see Table \ref{tab:relative_rate_by_sex_group}). The implementation of aging is a mechanism for people to enter and leave the sexually active population continuously and is necessary to propagate the effect of vaccination: we must ensure that vaccination of individuals in a particular age-group will later contribute to the number of vaccinated in older age-groups.   

Each of the equations (\ref{eq:gw_system_eq1})-(\ref{eq:gw_system_eq5}) describes the change in the number of individuals that occurs during a small time period as the sum of the number of individuals entering this compartment from other compartments and those leaving the compartment. Discussions on construction of compartmental disease transmission models are presented in \cite{Keeling2007} and \cite{Vynnycky2010}. Consider, for example, compartment $G$ (Figure \ref{fig:genital_warts_model}): we can calculate the change in the number of individuals in this compartment during a small interval of time by adding up the individuals entering the compartment during this time interval (the infected who developed genital warts, $\gamma_g I_{g,s,a}$ and the aging members of $G$ from age-group $a-1$, $G_{g,s,a}/5$ ) and subtracting the number that leave the compartment (recovered who go either to $P$ or $N$, $r_gG_{g,s,a}$ and the aging members of $G$ moving to age-group $a+1$, $G_{g,s,a}/5$). In so doing, we will obtain $\gamma_g I_{g,s,a}+G_{g,s,(a-1)}/5-r_gG_{g,s,a}-G_{g,s,a}/5$ which is the right-hand side of equation (\ref{eq:gw_system_eq3}).

It is necessary to point out the crucial role of the force of infection in our model (Table \ref{tab:parameter_interpretation}). This is the only non-constant coefficient we have to deal with, which introduces non-linearity into the system (\ref{eq:gw_system_eq1})-(\ref{eq:gw_system_eq5}). It depends on the age and risk-group specific patterns of sexual behaviour represented by a matrix usually known as a 'sexual mixing' matrix \cite{Anderson1992}.

In the following subsection we provide a concise description of the construction of the sexual mixing matrix. For complete details we refer the reader to Appendix 1 and an associated research paper \cite{Garnett1994}.

\subsection{Sexual Mixing Matrix}
\label{sec:smm}
In this section, we discuss the model choice, the construction and extensions we made in this paper to the development of a statistical model for the sexual interaction of members of the population, as defined by the sexual mixing matrix. We consider a Sexually Transmitted Infection (STI) transmission model which has many features in common with other STI models. For example the models for gonorrhea in \cite{Ghani1997} or those developed for HIV in \cite{Garnett1994} and \cite{Garnett1993}, which describe patterns of mixing between age and sexual activity-groups with respect to HIV in heterosexual communities. Like these other models, our approach relies on certain assumptions about the way individuals form their sexual partnerships. This partnership formation process is commonly referred to as "sexual mixing". A simplified model for sexual mixing can be described via a sexual mixing matrix as described in \cite{Anderson1992}, and for which examples can be found in \cite{Garnett1994}, \cite{Garnett1996}, \cite{Garnett1996a} and \cite{Ghani1998}.

It is not uncommon in the medical literature to assume the parameters of this mixing matrix to be known throughout the calibration of the resulting ODE epidemic model (see \cite{Garnett1996}). This is because these parameters are normally derived from the data obtained from an extensive sexual behaviour survey which often serves as the only comparatively reliable source of information on the subject. Therefore, if the number of participants in a particular survey is significantly larger than in other surveys, this survey will likely be considered as the most trustworthy source of information on sexual mixing, often irrespective of the experimental design and population studied. However, due to the personal nature of such surveys it is understood that their results are to be taken cautiously. In our model we will assume that two of the parameters specifying the sexual mixing matrix are unknown and  should be jointly estimated with the ODE model parameters.

In our heterosexual model formulation, the mixing matrix is a $\left(2\times4\times4\times9\times9\right)$ dimensional matrix comprised of the product terms $c_{g,s,s',a,a'}\rho_{g,s,s',a,a'}$, where  $c_{g,s,s',a,a'}$  is the mean per capita annual rate at which an individual of gender $g$ from a risk or activity-group $s$ and age-group $a$ acquires new sexual partners of the opposite gender $g'$ from a risk-group $s'$ and age-group $a'$; $\rho_{g,s,s',a,a'}$ is the conditional probability that an individual of gender $g$ from sexual activity-group $s$ and age-group $a$ acquires a sexual partner of the opposite gender $g'$ from sexual activity-group $s'$ and age-group $a'$. It is clear that estimation of all of these parameters is an almost insurmountable statistical challenge, which is one of the reasons why these parameters are often taken as fixed in any given calibration study of STI transmission models. There are two broad approaches one could pursue, the first, given we are working in a Bayesian modelling framework in this paper would involve prior elicitation for these population parameters based on expert opinions of annual interaction rates that would be reasonably understood by medical practitioners in sexual health clinics, sexual health workers and social workers in regions in which respondents were recorded. The other alternative involves re-parameterizing aspects of this matrix, simplifying it significantly, allowing one to account for the uncertainty associated with specification of this matrix in an appropriate simplified stochastic model. This would involve finding suitable factors common to aspects of this matrix that could instead be taken as stochastic and estimated in the model calibration, which in turn allow one to derive each element of the sexual mixing matrix. Most importantly, the framework we develop and present for the estimation and calibration of the transmission model is general enough to be used for either of these approaches and any degree of unknown parameters in the sexual mixing matrix and any parameterisation deemed suitable for a given population study.

In this paper, we then utilise this matrix to specify the force of infection (see equations (\ref{eq:gw_system_eq1}) and (\ref{eq:gw_system_eq2})) for any individual from any subgroup $(g,s,a)$ according to the parameterisation
\begin{equation}
\label{def:force_of_infection}
\lambda_{g,s,a}=\beta_g\sum_{s',a'}\left\{c_{g,s,s',a,a'}\rho_{g,s,s',a,a'} \frac{I_{g',s',a'}}{ S_{g',s',a'}+ I_{g',s',a'}+ G_{g',s',a'}+ P_{g',s',a'}+ N_{g',s',a'}}\right\},
\end{equation}
where $\beta_g$ is the transmission probability per partnership, i.e. the probability that a susceptible person of gender $g$ will become infected due to a partnership with an infected person of gender $g'$. As discussed in \cite{Garnett1994}, we can easily incorporate into specification of (\ref{def:force_of_infection}) several parameters allowing us to control to what extent the matrix should reflect assortative partnership formation patterns, i.e. how likely are individuals to find sexual partners among 'similar' individuals, thereby altering the properties of the matrix, making it more or less dense.
For example, we can vary the extent to which older males prefer to have younger female sexual partners, or, possibly, a tendency of older females to choose younger male sexual partners. It is also necessary to be able to adjust for or account for the readiness of each gender to compromise with the wishes of the opposite gender. This last point becomes significant whenever supply of individuals of one gender does not meet the demand for sexual partners from the individuals of the opposite gender.

In this paper, we will treat the degrees of assortativity by age and sexual activity-groups, observed to have a noticeable effect on the model calibration, as uncertain, while the rest of the sexual mixing matrix parameters will be fixed. We do not assume that all of the parameters specifying this matrix are unknown since we want to keep the total number of parameters in our model as low as possible.

While a matrix with these features cannot encompass all the complexities of human sexual mixing, it certainly enables us to explore various relatively plausible mixing scenarios. In addition it is a manageable model formulation and this framework has been widely used by STI modellers (for examples of its use in HPV models see \cite{Barnabas2006,Kim2007,Regan2007,Elbasha2007}).

\section{Bayesian framework for non-linear ODE HPV Epidemic model}
\label{BayesianModel}
In constructing a statistical model for the HPV epidemic model we treat the parameters of the non-linear system of ODE equations describing the epidemic as unknown random variables. In addition, we treat some of the parameters of the sexual mixing matrix associated with assortativity as unknown random variables to be estimated jointly with the ODE epidemic model parameters. We formulate a Bayesian model for this system with the prior specifications given in Table \ref{CoeffPrior}. Furthermore, we treat the ODE system as purely deterministic, conditional on a set of system parameters. In other words, the trajectories of the latent states in the ODE system over time are conditionally, deterministically specified by the system of ODEs. Therefore, we do not derive a system of stochastic differential equations as has been done in the Pharmaco-Kinetic/Pharmaco-dynamics literature (see \cite{DitlevsenDegaetano2005mixed} or \cite{DitlevsenDegaetano2005rat}). Instead our focus lies purely in the "calibration" of this ODE epidemic model to an observed set of data based on observations which are formed by a transformation of the state of the ODE system and observed in noise. We detail the process of forward simulation we utilise to obtain the state of the non-linear ODE system at any given time point, conditional on the model parameters defined in Section \ref{ForwardSim}.

\subsection{Forward simulation of the non-linear ODE model}
\label{ForwardSim}
The states of the system at time points $t \in \left\{1,\ldots,T\right\}$ are denoted by vectors $\bm{X}_{g,a,s}(1:T) = \left[\bm{X}_{g,a,s}(1),\ldots,\bm{X}_{g,a,s}(T)\right],$ which generically represent the dynamic evolution of the ODE system, where for a given gender, age and sexual activity-group at time $t$ the state encompasses the following components $\bm{X}_{g,a,s}(t) = \left[S_{g,a,s}(t),I_{g,a,s}(t),G_{g,a,s}(t),P_{g,a,s}(t),N_{g,a,s}(t)\right]$. Therefore, conditional on a set of parameters in the ODE system, we can iterate the ODE system forward in time using an ODE solver to obtain the states for each population group at time $t$. Note, the times $t$, for which the system is solved, will generally be of a much finer granularity to those at which observations are collected in a population study. This set of times should include, for the minimum subset of times for which the system is solved, the observation times.

In our estimation procedure we considered several different solvers, and noticed that the choice of solver can have a significant effect on the accuracy and on the efficiency of the computations undertaken in the statistical estimation. In particular, depending on the parametrization and development of the relationships between states or compartments of the model and the parametrization of the mixing matrix, one can obtain non-stiff or stiff systems of equations. In such settings repeated application of generic solvers will be computationally inefficient.

The software implementation of the framework described in the paper was written in Fortran 90/95/2003 and utilised a universal solver \texttt{dodesol} which is a part of the Intel$^{\textsc{\textregistered}}$ Ordinary Differential Equation Solver Library. This solver is, in fact, a collection of five different solvers designed to solve numerically initial value problems of the form
\begin{equation}
\label{eq:general_form_for_ODE_solver}
\dot{\bm{x}} = \bm{g}(\bm{x},t), \, \, \bm{x}(t_0) = \bm{x}_0,\,\,t>t_0,
\end{equation}
where $t$ is an independent variable, $\bm{x}$ is a vector of state variables to be solved for and
$g(\bm{x},t)$ is a function of $t$ and $\bm{x}$.

An important distinction between these solvers is that each of them being of explicit, implicit or hybrid type is particularly  efficient for a given degree of stiffness of the system. The universal solver first estimates how stiff the system is and then employs an appropriate solver to handle it. It is pertinent to note that in the case where (\ref{eq:general_form_for_ODE_solver}) can be solved explicitly a modification of the fourth order Merson's method combined with a first order multi-stage method will be applied. Should  (\ref{eq:general_form_for_ODE_solver}) be treated implicitly, an $L$-stable fourth order $(5,2)$-method with an option to fix the Jacobian will be used.

We chose the \texttt{dodesol} solver not only because of its convenient Fortran interface, for which the remainder of the model and code was developed, but also  efficient performance which we compared with that of the two standard MATLAB solvers \texttt{ode45} and \texttt{ode15s}. The solver \texttt{dodesol} was especially superior to those provided by MATLAB solvers \texttt{ode45} and \texttt{ode15s} when the stiffness of our ODE system increased. However, it should be pointed out that \texttt{dodesol} speed can be considerably affected by the selection of its minimal time step and relative error threshold. To obtain the simulation results discussed in this paper we used the minimal step size $1\times 10^{-10}$, the initial step size $1\times 10^{-6}$ and the relative error tolerance $1\times 10^{-3}$. Also, we did not make use of the option to pre-specify the Jacobian provided by the solver.

\subsection{Prior choices for parameters of non-linear ODEs and the sexual mixing matrix}
\label{PriorSection}
In this section we specify prior model developed for the non-linear ODE system modelling the epidemic of HPV as well as the extension to the  mixing matrix, making it stochastic. We summarise the prior choices made for each of the model parameters below in Table \ref{CoeffPrior}, where we have worked with a posterior parameterisation based on the input components of the model parameters in Table \ref{tab:parameter_interpretation}. Combining these prior choices, we can present the following basic prior structure:
\begin{multline}
%\begin{split}
p(TRm,TRf,WIPm, WIPf, DWTm, DWTf, DAIm, DAIf, PSCm, PSCf,\\ DIm, DIf,\Sigma,A_Y,EPSa,EPSr)
= p(TRm)p(TRf)p(WIPm)p(WIPf)p\times \\
(DWTm)p(DWTf)p(DAIm)p(DAIf)p(PSCm)p(PSCf)\times \\p(DIm)p(DIf)p(\Sigma)p(A_Y)p(EPSa) p(EPSr).
%\end{split}
\label{PriorMod}
\end{multline}

We also note, that when specifying the prior distributions and their hyper-parameters (i.e. parameters describing the shape of distributions assigned to the prior distribution parameters) we utilise data reported in the literature from previous studies of HPV transmission to inform some appropriate prior specifications (see Table \ref{tab:real_data}). Importantly, these data are completely different to the real data sets that were studied in the present paper, based on different populations, different time periods and different locations, hence we have not used our observation data twice. Furthermore, where data were not informative on particular parameters, uninformative prior specifications were utilised. As is usually the case with interpreting any reported HPV data, it is appropriate to consider the reported values of observations and calibrations as nothing more than an estimation or a trial with its own particular limitations and proneness to error. This means we can safely consider values outside of the reported confidence intervals to be plausible in our study.
Using (\ref{tab:real_data}) we specify the priors for each of the HPV models (Table \ref{tab:used_priors}).

The prior parameters for seroprevalence observation errors denoted by $A_Y$ had hyper-prior parameters given by $(k_{Am},\theta_{Am}) =(2,2)$ and the diagonal element $\sigma$ of the incidence observation error covariance matrix was distributed according to $(k_{\sigma},\theta_{\sigma})=(2,5)$. Finally, the priors for the sexual mixing matrix parameters associated with assortativity had hyper-priors specified according to $\left(\alpha_{\epsilon_a},\beta_{\epsilon_a}\right) = \left(0.5,0.7 \right)$ and $\left(\alpha_{\epsilon_r},\beta_{\epsilon_r}\right) = \left(0.5,0.7 \right)$.

\subsection{Likelihood model for parameters of non-linear ODEs and Garnett mixing matrix}
\label{LikelihoodSection}
Having specified the prior structure of the model we present the observation model, likelihood and details of the actual data studied. In particular, we note that conditional on a particular realisation of the model parameters in Table \ref{CoeffPrior}, the latent unobserved state trajectories of the system are deterministic, non-analytic solutions to the system of ODEs in Equations (\ref{eq:gw_system_eq5}). These solution trajectories are transformed according to age, gender and risk-group to produce a set of results that can be directly observed in noise in the population under study. In the following section we will describe and detail these transformations and the associated statistical assumptions made on the observation error. In doing so we will also specify models suitable for situations of disequilibrium of the disease states within the population such as occurs after a serious outbreak, a vaccination or community education awareness program, etc., and models for situations in which the disease in the population under study can be considered to be at equilibrium.

In this section we first describe the observations that are utilised in the calibration of the ODE epidemic model and the sexual mixing matrix. Importantly, the real observations utilised in this paper, based on HPV-6/-11 data collected in Australia, will be treated as taken from a population at endemic equilibrium.

Then we present the resulting likelihood model utilised, followed by derivation of the posterior distribution. We first note that the general likelihood model will be presented in which observations of the transformed state trajectories are known at a fixed set of discrete time points $t \in \left\{1,2,\ldots,T\right\}$. In practice, these discrete time points may be unevenly spaced and this would not affect the estimation procedure to be developed.

The actual observations we consider in this model are seroprevalence and genital warts incidence by age-group, which make up the "observations" considered in our model. They are specified as below:
\begin{description}
\item[Seroprevalence] is the proportion (or percentage, in our model) of individuals in a given  group who are recovered and seropositive;
\item[Genital warts incidence] is a rate defined as the number of new cases of genital warts in the population per person (or per 1000 persons, in our model) per year.
\end{description}
In order to evaluate the output of our model, we intend to compare it with the real-life seroprevalence \cite{Newall2008} and genital warts incidence \cite{Pirotta2009} data. The seroprevalence data were collected in the second half of 2005 from various laboratories in New South Wales, Victoria and Queensland where about 80\% of the Australian population resides. A validated sampling method for serosurveillance was applied to test 1523 serum samples from females and 1247 samples from  males aged from 0 to 69 years. For each individual the age-group, sex, and date of sample collection were recorded. The overall population HPV seroprevalence was derived by weighting the results obtained from the samples to 2005 Australian midyear population estimates by age. Incidence data were estimated from the Bettering the Evaluation of Care and Health (BEACH) cross sectional database. In particular, the annualised new case consultation rate stratified by gender and age from April 2000 to September 2006  was analysed and normalised to the corresponding 2004 Australian population. In total, 639  consultations related to genital warts were registered and of these roughly 35\% were new cases. BEACH did not capture the people who attended sexual health clinics to seek treatment for genital warts. However, it was estimated that for every person visiting a GP there was 0.298 persons visiting a sexual health clinic, so the genital warts cases managed in such clinics were accounted for by multiplying the incidence rates obtained from the GP data by 1.298.
The data are summarised in Tables \ref{tab:seroprevalence_AU} and \ref{tab:GW_incidence_AU}.

Generally, observations can be considered in this model as arriving at irregularly spaced intervals and generally not at all times $t \in \left\{1,\ldots,T\right\}$. For simplicity, we will assume that a full panel of observations for all age-groups, activity-groups and genders is available at each observation time, though this assumption can also be relaxed under the model developed in this paper. We will also assume for simplicity that since we can solve for the state of the non-linear system at any specified time point, conditional on a set of model parameters. We will always have the observation time corresponding to one of the time points $t \in \left\{1,\ldots,T\right\}$.

We model the observation vector as the result of a vector function of the non-linear ODE system state $\bm{X}_{g,a,s,t}$ at each time $t$, denoted by $\bm{O}_{g,a,s}(t) = \left[D_{g,a,s}(t), Y_{g,a,s}(t)\right]$ which contains the observation $D_{g,a,s}(t)$ representing incidence at time $t$ for a given gender, age-group and sexual activity-group as well as $Y_{g,a,s}(t),$ which represents the seroprevalence in the given category.

The observed numbers of new diagnoses are assumed to be observed in Gaussian noise. This assumption is reasonable since the counts obtained are very large, so it is suitable to make a continuous distributional assumption. The likelihood model for the observed seroprevalences is assumed to be a beta distribution since it represents observations of proportions given the parameters. Furthermore, conditional on the parameters of the model, the joint likelihood model is assumed to have the following conditional independence properties between observations conditional on information on the individuals categories of gender, age and activity-group as follows:
\begin{multline}
\mathcal{L}\left(\textbf{TR}, \textbf{WIP},\textbf{GWT}, \textbf{DAI}, \textbf{PSC}, \textbf{DI}, \Sigma, A_Y, EPSa, EPSr,  \bm{X}_{g,a,s}(1), \ldots, \bm{X}_{g,a,s}(T);\right. \\
\left. \bm{O}_{g,a,s}(1),\ldots,\bm{O}_{g,a,s}(T)\right)= \prod_{g\in\left\{1,2\right\}}\prod_{a\in\left\{1,\ldots,9\right\}}\prod_{s\in\left\{1,\ldots,4\right\}} \prod_{t=1}^TP\left(\bm{O}_{g,a,s}(t)| \textbf{TR}, \textbf{WIP},\textbf{GWT},\right.\\
\left. \textbf{DAI}, \textbf{PSC}, \textbf{DI}, \Sigma, A_Y, EPSa, EPSr, \bm{X}_{g,a,s}(t)\right)= \prod_{g\in\left\{1,2\right\}}\prod_{a\in\left\{1,\ldots,9\right\}}\prod_{s\in\left\{1,\ldots,4\right\}} \prod_{t=1}^T P(D_{g,a,s}(t)| \textbf{TR},\\
\textbf{WIP},\textbf{GWT}, \textbf{DAI}, \textbf{PSC}, \textbf{DI}, \Sigma, EPSa, EPSr, \bm{X}_{g,a,s}(t))\times\\
 P\left(Y_{g,a,s}(t)| \textbf{TR}, \textbf{WIP},\textbf{GWT}, \textbf{DAI}, \textbf{PSC}, \textbf{DI}, A_Y, EPSa, EPSr,  \bm{X}_{g,a,s}(t)\right)\\
= \prod_{g\in\left\{1,2\right\}}\prod_{a\in\left\{1,\ldots,9\right\}}\prod_{s\in\left\{1,\ldots,4\right\}} \prod_{t=1}^T \mathcal{N}\left(D_{g,a,s}(t); \bm{\mu}_{D_{g,a,s}(t)}, \sigma_{D_{g,a,s}(t)}\right) Be\left(Y_{g,a,s}(t); A_Y,B_Y\right),
\label{LHModel}
\end{multline}
where we denote in bold the vectors of parameters corresponding to males and females. For example, $\bm{TR}=(TRm,TRf)$; the equilibrium mean structure for the number of new genital warts diagnoses (incidence) of HPV-6 and HPV-11 in category $g,a,s$ is defined as
\begin{equation}
\label{eq:incidence_at_equilibrium}
\begin{split}
\bm{\mu}_{D_{g,a,s}(t)} &= f\left(\textbf{TR}, \textbf{WIP},\textbf{GWT}, \textbf{DAI}, \textbf{PSC}, \textbf{DI}, \Sigma, EPSa, EPSr, \bm{X}_{g,a,s}(t)\right) \\
&= \frac{1}{WIP_g} \times \frac{I_{g,s,a}}{S_{g,s,a}+I_{g,s,a}+G_{g,s,a}+P_{g,s,a}+N_{g,s,a}} \times 1000,
\end{split}
\end{equation}
which is a function of the state vector at time $t$ and the model parameters. The multiplier 1000 appears here because  we compare the simulated incidence with the real incidence data reported per 1000 individuals \cite{Pirotta2009}.

In this paper we always consider gender to be male or female coded with a 1 or 2 respectively, sexual activity to be split into four classes coded with 1,2,3,4 and ages as decomposed in Table \ref{tab:relative_rate_by_age_group} coded by the indices 1 through 9.
It has to be stressed that (\ref{eq:incidence_at_equilibrium}) is valid only under the assumption that our model is at endemic equilibrium. Otherwise, the incidence would be time dependent: for any two time points $t_0$ and $t_1$ such that $t_1=t_0+1$ (they are 1 year apart)  would be given by the following:
\begin{equation}
\label{eq:incidence_not_at_equilibrium}
\bm{\mu}_{D_{g,a,s}(t_1)}= \frac{1}{WIP_g} \times \frac{I_{g,s,a}(t_0)}{S_{g,s,a}(t_0)+I_{g,s,a}(t_0)+G_{g,s,a}(t_0)+P_{g,s,a}(t_0)+N_{g,s,a}(t_0)} \times 1000.
\end{equation}

Note, we have obtained the state vectors for the system $\bm{X}_{g,a,s}(t)$ used in the evaluation of the likelihood model via forward simulation, conditional on the model parameters of the non-linear ODE model, ensuring that the solution points at which we evaluated the states of the system at least corresponded to the observation times $t \in \left\{1,\ldots,T\right\}$ and generally would be a finer grid than these time points. This was achieved using a specialised ODE solver as described previously in Section \ref{ForwardSim}.

Additionally, we also define the observation equation for the likelihood corresponding to the seroprevalence observations for each age-group, activity-group and gender, as a function of the non-linear ODE system state and model parameters, which were specified according to a Beta distribution specified in Equation (\ref{SPLH}) below:
\begin{equation}
\begin{split}
Y_{g,a,s}(t) &= g\left( \textbf{TR},\textbf{WIP},\textbf{GWT}, \textbf{DAI}, \textbf{PSC}, \textbf{DI}, A_Y, EPSa, EPSr, \bm{X}_{g,a,s}(t)\right)\\
&= \frac{P_{g,s,a}}{S_{g,s,a}+I_{g,s,a}+G_{g,s,a}+P_{g,s,a}+N_{g,s,a}}.
\end{split}
\label{SPLH}
\end{equation}

The scale parameter $A_Y$ of this Beta prior is treated as a random variable and estimated in the posterior inference. The shape parameter, denoted by $B_Y$ is obtained by moment matching given the sampled scale parameter and the observations obtained from the current forward projection trajectory:
\begin{equation}
\begin{split}
B_Y = A_Y \left(\frac{1}{Y_{g,a,s}(t)}-1\right).
\end{split}
\label{B_Y}
\end{equation}

We may then combine this likelihood and prior structure to obtain the posterior distribution of the model parameters, given the observations. The full posterior distribution of interest is given in Equation (\ref{Post}) and is comprised of the likelihood model in Equation (\ref{LHModel}) and the prior model in Equation (\ref{PriorMod}):
\begin{multline}
p\left(\textbf{TR}, \textbf{WIP},\textbf{GWT}, \textbf{DAI}, \textbf{PSC}, \textbf{DI}, \Sigma, A_Y, EPSa, EPSr| \bm{O}_{g,a,s}(1),\ldots,\bm{O}_{g,a,s}(T)\right) \\
\propto 
\mathcal{L}\left( \textbf{TR}, \textbf{WIP},\textbf{GWT}, \textbf{DAI}, \textbf{PSC}, \textbf{DI}, \Sigma, A_Y, EPSa, EPSr, \bm{X}_{g,a,s}(1), \ldots, \bm{X}_{g,a,s}(T); \right.\\
\left. \bm{O}_{g,a,s}(1),\ldots,\bm{O}_{g,a,s}(T)\right)p(TRm)p(TRf)p(WIPm)p(WIPf)p(DWTm)p(DWTf)p\times\\
(DAIm)p(DAIf)p(PSCm)p(PSCf) p(DIm)p(DIf)p(\Sigma)p(A_Y)p(EPSa) p(EPSr).
\label{Post}
\end{multline}

The resulting posterior distribution therefore has 16 model parameters. At each time point $t$, the forward simulated ODE state vectors $\bm{X}(t) \in \mathbb{R}^{360}$, due to 2 sexes, 9 age-groups, 4 activity-groups and 5 compartments. Hence, the total dimension explored in the simulation performed in the results section is $360 \times T$, where we set $T=120$.

Next we demonstrate how to perform inference under this Bayesian model formulation. In particular we are interested in the Bayesian point estimators corresponding to the posterior mean (MMSE) and the posterior mode (MAP) estimates, as well as the distribution properties of the posterior. To explore these we must resort to a numerical procedure to draw samples from the posterior distribution. The class of methods we consider in this paper involves combining the forward simulation procedure for solving the non-linear ODE system given a set of model parameters, with an adaptive Markov chain Monte Carlo (AdMCMC) algorithm to propose a new set of model parameters given past history of proposed parameter vectors. The details of the AdMCMC methodologies considered are presented in Section \ref{AdMCMCMethodology}.

\section{Adaptive MCMC Strategies Combined with Forward Simulation}
\label{AdMCMCMethodology}
In this section we present details for the adaptive Markov chain Monte Carlo (AdMCMC) algorithm that will be combined with the Forward simulation algorithm in order to sample from the posterior distribution given in Equation (\ref{Post}). In particular, we first introduce the background of an AdMCMC algorithm, before detailing the adaptation strategy we will explore in this paper based on the adaptation rules developed in \cite{Atchade2005, Roberts2009}.

In essence, we first construct a MCMC proposal distribution to sample the static posterior parameters, denoted by vector $\bm{\Theta} = \left[ \textbf{TR}, \textbf{WIP},\textbf{GWT}, \textbf{DAI}, \textbf{PSC}, \textbf{DI}, \Sigma, A_Y, EPSa, EPSr\right]$. Then, conditional on these model parameters proposed, we obtain the state trajectories for the ODE HPV epidemic model, given by $\bm{X}_{g,a,s}(1:T) = \left[\bm{X}_{g,a,s}(1),\ldots,\bm{X}_{g,a,s}(T)\right]$, which are generated using forward simulation. The MCMC proposal distribution is parameterised by parameter vector $\bm{\Phi}$. The idea of adaptive MCMC methods is to learn appropriate values for $\bm{\Phi}$ recursively utilising the previous samples of the Markov chain that have been accepted under the MCMC accept-reject mechanism. This is achieved on-line, adapting according to the support of the posterior distribution, thereby allowing the Markov chain to discover and explore the regions of the posterior distribution that have the most mass. Through this on-line adaptive learning mechanism the Markov chain proposal distribution can significantly improve the acceptance rate of the Markov chain, enabling efficient mixing and improving the samples obtained from the posterior.

In this paper we consider a popular adaptation scheme proposed in the literature and analyse its ability to explore the very high dimension of the posterior distribution developed in the non-linear HPV epidemic model. Before presenting the details of the adaptation scheme for the Markov proposal, we first present the generic algorithm developed for sampling from the posterior in Equation (\ref{Post}). In the following sequence of steps for the $j$-th iteration of the AdPMCMC Forward Simulation algorithm, we will update the state of the Markov chain from $\bm{\Theta}^{(j-1)}$, with corresponding states $\bm{X}^{(j-1)}_{g,a,s}(1:T)$, to parameter vector $\bm{\Theta}^{(j)}$ with associated state trajectories $\bm{X}^{(j)}_{g,a,s}(1:T)$. This algorithm is summarised below for one step of the AdMCMC Forward project algorithm:
\begin{enumerate}
\item{Sample $\bm{\theta}^{*} \sim q\left([\bm{\theta}](j-1),\cdot\right)$ from an Adaptive MCMC proposal constructed using previous Markov chain samples $\left\{\bm{\Theta}^{(1)},\ldots,\bm{\Theta}^{(j-1)}\right\}$.}
\item{Solve the non-linear ODE system (\ref{eq:gw_system_eq1})-(\ref{eq:gw_system_eq5}), by running a forward simulation of the non-linear ODE solver, conditional on proposed parameter vector $\bm{\theta}^{*}$, to obtain the state of the ODE system, $\bm{X}_{g,a,s}(1),\ldots,\bm{X}_{g,a,s}(T)$, which correspond to the observations \\
$\bm{O}_{g,a,s}(1),\ldots,\bm{O}_{g,a,s}(T)$.}
\item{Accept the proposed new Markov chain state comprised of $\bm{\theta}^{*}$ with acceptance probability given by
%{\small{
\begin{equation*}
\alpha\left(\bm{\theta}^{(j-1)},\bm{\theta}^{*}\right)
 = \min \left( 1, \frac{\mathcal{L}\left(\bm{\theta}^{*}; \bm{O}_{g,a,s}(1),\ldots,\bm{O}_{g,a,s}(T)\right) p\left(\bm{\theta}^{*}\right) q\left(\bm{\theta}^{(j-1)},\bm{\theta}^{*}\right)} {\mathcal{L}\left(\bm{\theta}^{(j-1)}; \bm{O}_{g,a,s}(1),\ldots,\bm{O}_{g,a,s}(T)\right) p\left(\bm{\theta}^{(j-1)}\right) q\left(\bm{\theta}^{*},\bm{\theta}^{(j-1)}\right)} \right)
\end{equation*}
%}} 
where we evaluate this acceptance probability utilising the expressions developed in Section \ref{PriorSection}, Section \ref{PriorSection} and Section \ref{AdaptionSection}. These steps are repeated for $j \in \left\{1,\ldots,J\right\}$.
}
\end{enumerate}
\vspace{0.5cm}

\subsection{Adaptive Metropolis}
\label{AdaptionSection}
To complete the specification of the methodology utilised, we present the internal adaptation strategy we considered in this paper based on the adaptive Metropolis algorithm detailed in \cite{Roberts2009}. This is a variant of the approach proposed in \cite{Haario2001} which develops a Random Walk Metropolis-Hastings (RWMH) that estimates the global covariance structure from the past samples.

Under an Adaptive Metropolis algorithm, the proposal distribution is based on a Gaussian mixture kernel detailed in \cite{Roberts2009}. The proposal, $q\left(\bm{\theta}^{(j-1)},\bm{\theta}^{*}\right)$, involves an adaptive Gaussian-mixture Metropolis proposal, one component of which has a covariance structure that is adaptively learnt on-line as the algorithm explores the posterior distribution. For iteration $j$ of the Markov chain the proposal is
\begin{equation}
q_j\left(\bm{\theta}^{(j-1)},\cdot\right)= \gamma
\mathcal{N}\left(\bm{\theta}^{*};\bm{\theta}^{(j-1)},\frac{\left(2.38\right)^2}{d}\Sigma_j\right)
+ \left(1-\gamma\right)
\mathcal{N}\left(\bm{\theta}^{*};\bm{\theta}^{(j-1)},\frac{\left(0.1\right)^2}{d}I_{d,d}\right).
\label{AdaptiveProp}
\end{equation}
Here, $\bm{\Phi} = \Sigma_j$ is the current empirical estimate of the covariance between the parameters of $\bm{\theta},$ estimated using samples from the Markov chain up to time $j-1$. The theoretical motivation for the choices of scale factors 2.38, 0.1 and dimension d are all provided in \cite{Roberts2009} and are based on optimality conditions presented in \cite{Roberts2001}. We note that the update of the covariance matrix, can be done recursively on-line via the following recursion (as detailed in \cite{Atchade2009}):
\begin{equation}
\begin{split}
\mu_{j+1} &= \mu_{j} + \frac{1}{j+1}\left(\tilde{\beta}^{(j-1)} - \mu_j\right)\\
\Sigma_{j+1} &= \Sigma_{j} + \frac{1}{j+1}\left(\left(\tilde{\beta}^{(j-1)} - \mu_j\right)\left(\tilde{\beta}^{(j-1)} - \mu_j\right)' - \Sigma_j\right).
\end{split}
\end{equation}

\section{Synthetic Data Analysis}
In this section we first study the performance of the AdMCMC Forward simulation methodology in a synthetic data example. To perform this study we utilised model parameters randomly selected as: 
\begin{multline}
\left(TRm,TRf, WIPm, WIPf, DWTm, DWTf, DAIm, DAIf, PSCm, PSCf,\right. \\
\left. \Sigma, A_Y, EPSa, EPSr \right) =
\left(0.9, 0.9, 0.95, 0.85, 0.15, 0.3,  3.2,  3.4,  0.5,  0.6,   5.0,   2.0, 0.5,        0.5 \right).
\end{multline}
and distributed as shown in Table \ref{tab:used_priors}.
Note that $DIm$ and $DIf$ are omitted because we decided to assume a life-long immunity for all individuals in the population. This simplification can easily be removed and would not modify our estimation procedure or models developed.

With an initial population size of 10,000 we simulated trajectories of system (\ref{eq:gw_system_eq1})-(\ref{eq:gw_system_eq5}) over annual time steps for 120 years ($t\in [0,120]$) and every 10 years we calculated the synthetically generated "true" state and a noisy set of observations for each age-group under the specified statistical models in Equation \ref{LHModel}. These ODE state trajectories and the  observations were taken as the "true" synthetic data. Next we ran 100,000 iterations of the AdMCMC Forward Projection sampler to obtain samples from the posterior distribution conditional on these synthetically generated observations. 

\begin{figure}[h!]
\begin{center}
\includegraphics[scale=0.5]{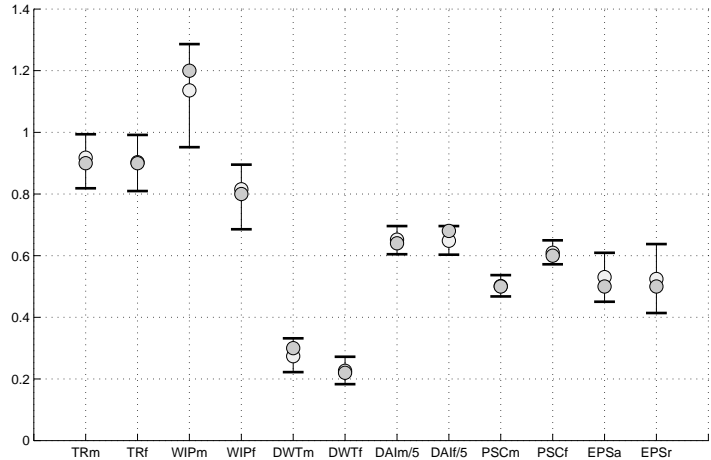}
\caption{"True" model parameters (gray circles) for synthetic study versus estimated model parameters (light gray circles) and error bars (95\% posterior C.I.) for the model corresponding to the combined HPV-6 and -11 case.}
\label{fig:parameters_synth}
\end{center}
\end{figure}

\begin{figure}[h!]
\begin{center}
\includegraphics[scale=0.6]{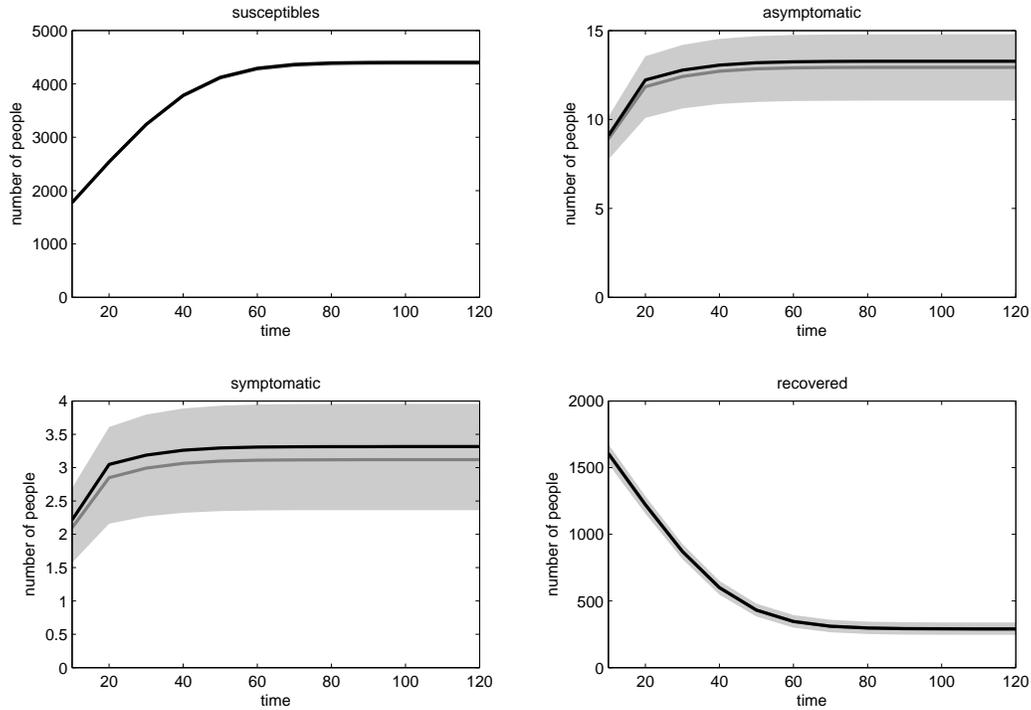}
\caption{State trajectories for males in every disease state (synthetic case, combined HPV-6 and -11): "true" trajectories (black line), MMSE (gray line) and 95\% CI (shaded area); note that in the presented case the trajectories for seropositive and seronegative males coincide, so it is convenient to call males in either state "recovered".}
\label{fig:states_synth_males}
\end{center}
\end{figure}

In Figure \ref{fig:parameters_synth} we present the posterior estimates for the ODE epidemic static model parameters (calibration performance) under the Bayesian model constructed in Section \ref{BayesianModel}. We illustrate this performance for the model parameters that describe the rates of transmission between states as well as those that quantify the statistical uncertainty we model in the sexual mixing matrix. Each of these parameters is estimated based on observations generated for the number of new diagnoses (incidence) and seroprevalence utilising prior specifications derived from previous literature studies. The important features of this synthetic study is that we can illustrate that our Adaptive MCMC Forward Projection sampling methodology is able to generate samples from the resulting high dimensional posterior efficiently and that the estimated posterior parameter MMSE values and 95\% posterior C.I. contain, in all cases, the "true" model parameters utilised to generate the data. These results demonstrate that the AdMCMC Forward project estimation procedure we developed is working accurately in this controlled synthetic study. This is also confirmed by the estimation of state trajectories. An example of estimated state trajectories for the total number of males in every disease state is shown in Figure \ref{fig:states_synth_males}. For all simulated time steps and for each disease state we observe that the "true" trajectory is contained within the 95\% posterior confidence interval for the aggregated trajectory and, in addition, the estimated mean trajectory is in agreement with the "true" aggregated trajectory.

\begin{figure}[h!]
\begin{center}
\includegraphics[width=\textwidth, height=0.4\textheight]{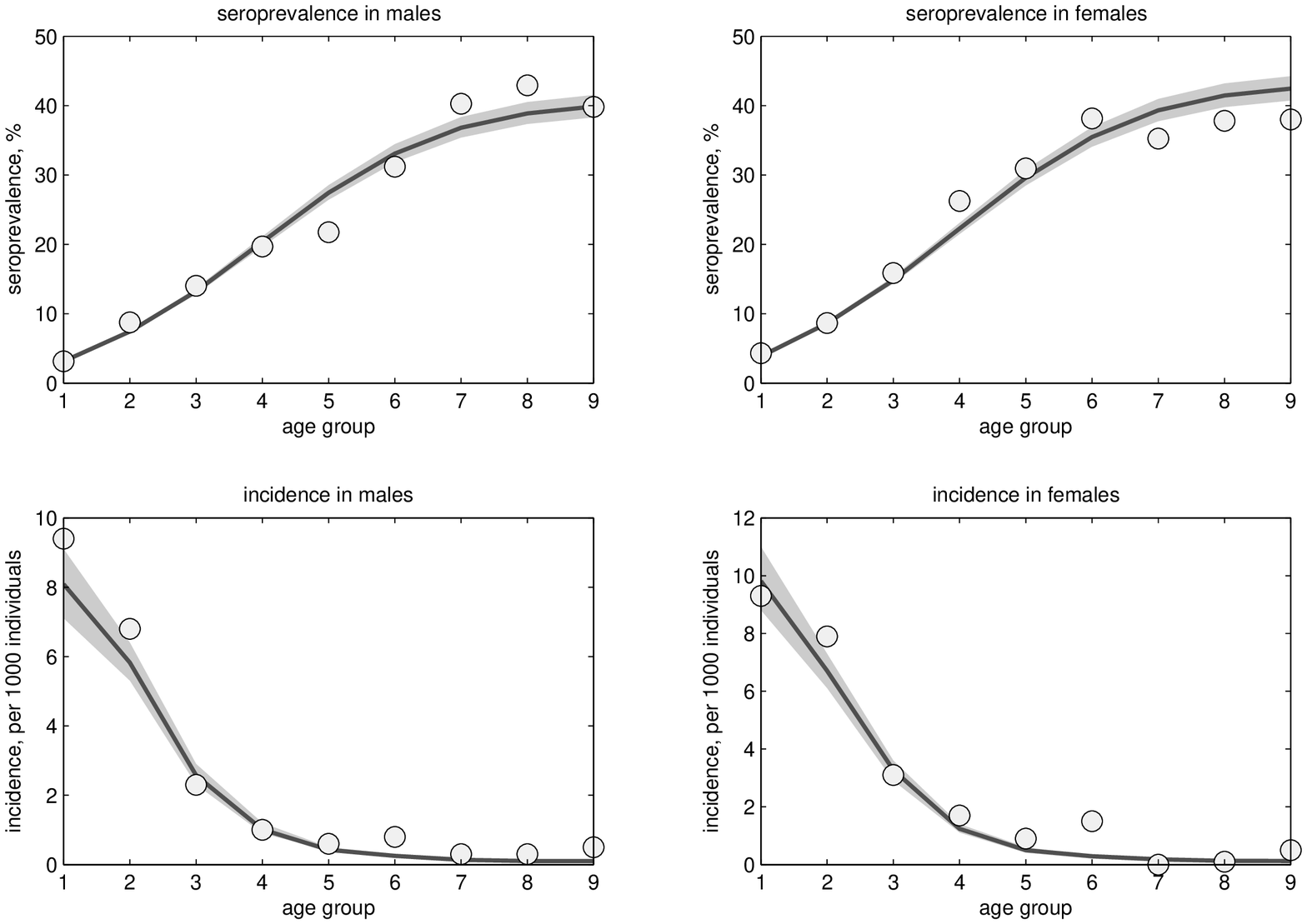}
\caption{Calibration for synthetic case at time point $t=30$ (combined HPV-6 and -11): MMSE (solid line) and 95\% CI (shaded area).}
\label{fig:calibration_synth_tp3}
\end{center}
\end{figure}

\begin{figure}[h!]
\begin{center}
\includegraphics[width=\textwidth, height=0.45\textheight]{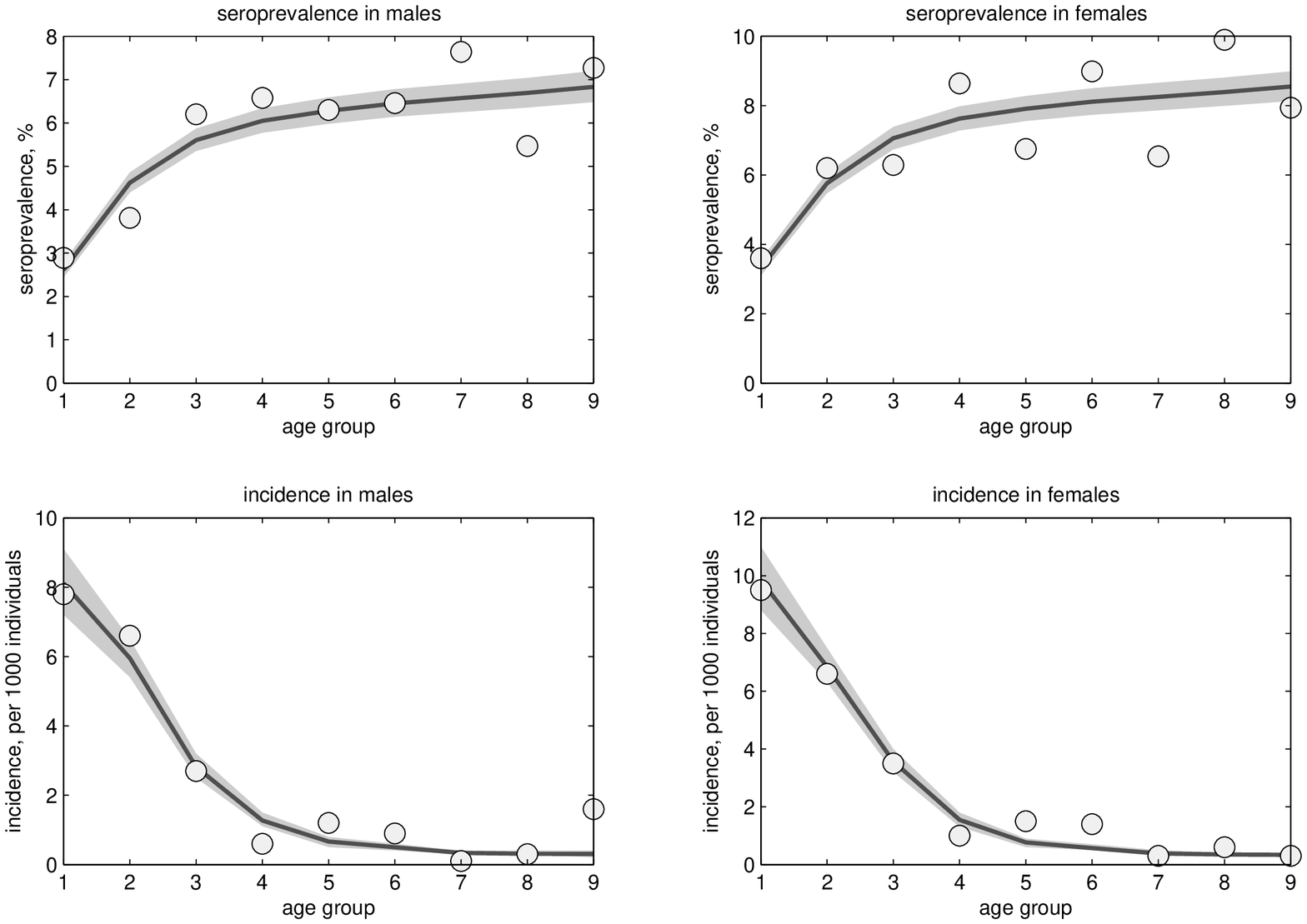}
\caption{Calibration for synthetic case at time point $t=120$ (combined HPV-6 and -11): MMSE (solid line) and 95\% CI (shaded area).}
\label{fig:calibration_synth_tp11}
\end{center}
\end{figure}

Furthermore, we can also provide analysis of the calibration performance as a function of the observed data versus estimated model fits to incidence and seroprevalence per age-group. The 95\% posterior confidence intervals of the simulated forecast posterior calibrations are compared to and found in good agreement with the "true" synthetically generated observations, as depicted in Figures \ref{fig:calibration_synth_tp3} and \ref{fig:calibration_synth_tp11}, for the number of new genital warts diagnoses and the seroprevalence by age and gender. 


\section{Real Data Analysis and Results}
In this section we undertook statistical estimation and performed calibration of the ODE model to the real data observations for HPV-6, HPV-11 and HPV-6 and -11 combined. We maintained the assumption that the duration of immunity is life-long for all individuals, since we found that as immunity was chosen closer to permanent the calibration became consistently better. This was measured with respect to the predictive performance of the posterior MMSE and associated posterior 95\% C.I. obtained from the model for each age-group and gender for seroprevalence and number of new diagnoses (incidence).

\begin{figure}[h!]
\begin{center}
\includegraphics[scale = 0.6]{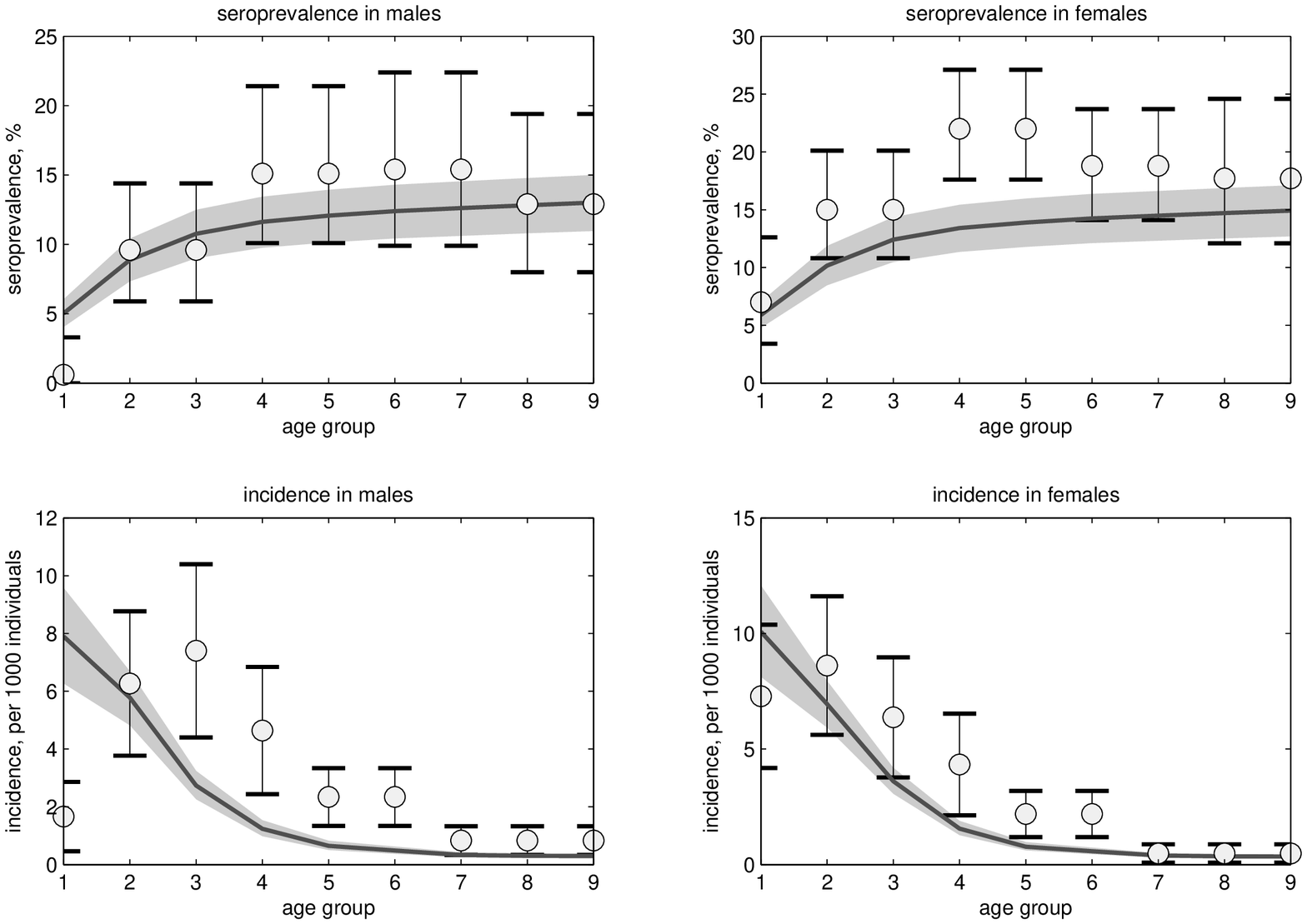}
\caption{Calibration for HPV-6: MMSE (solid line) and 95\% CI (shaded area).}
\label{fig:calibration_hpv6}
\end{center}
\end{figure}

\begin{figure}[h!]
\begin{center}
\includegraphics[scale = 0.6]{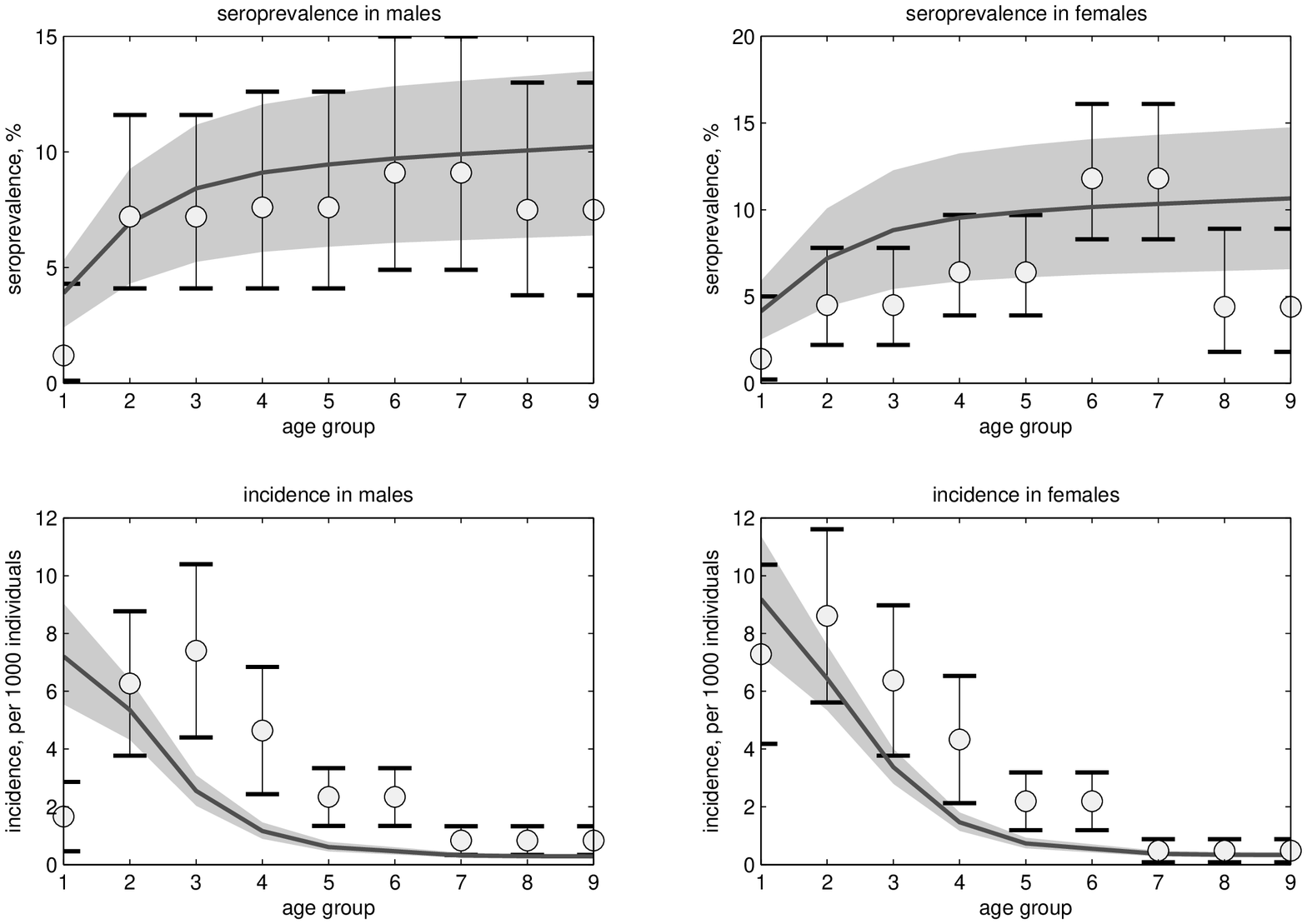}
\caption{Calibration for HPV-11: MMSE (solid line) and 95\% CI (shaded area).}
\label{fig:calibration_hpv11}
\end{center}
\end{figure}

\begin{figure}[h!]
\begin{center}
\includegraphics[scale = 0.6]{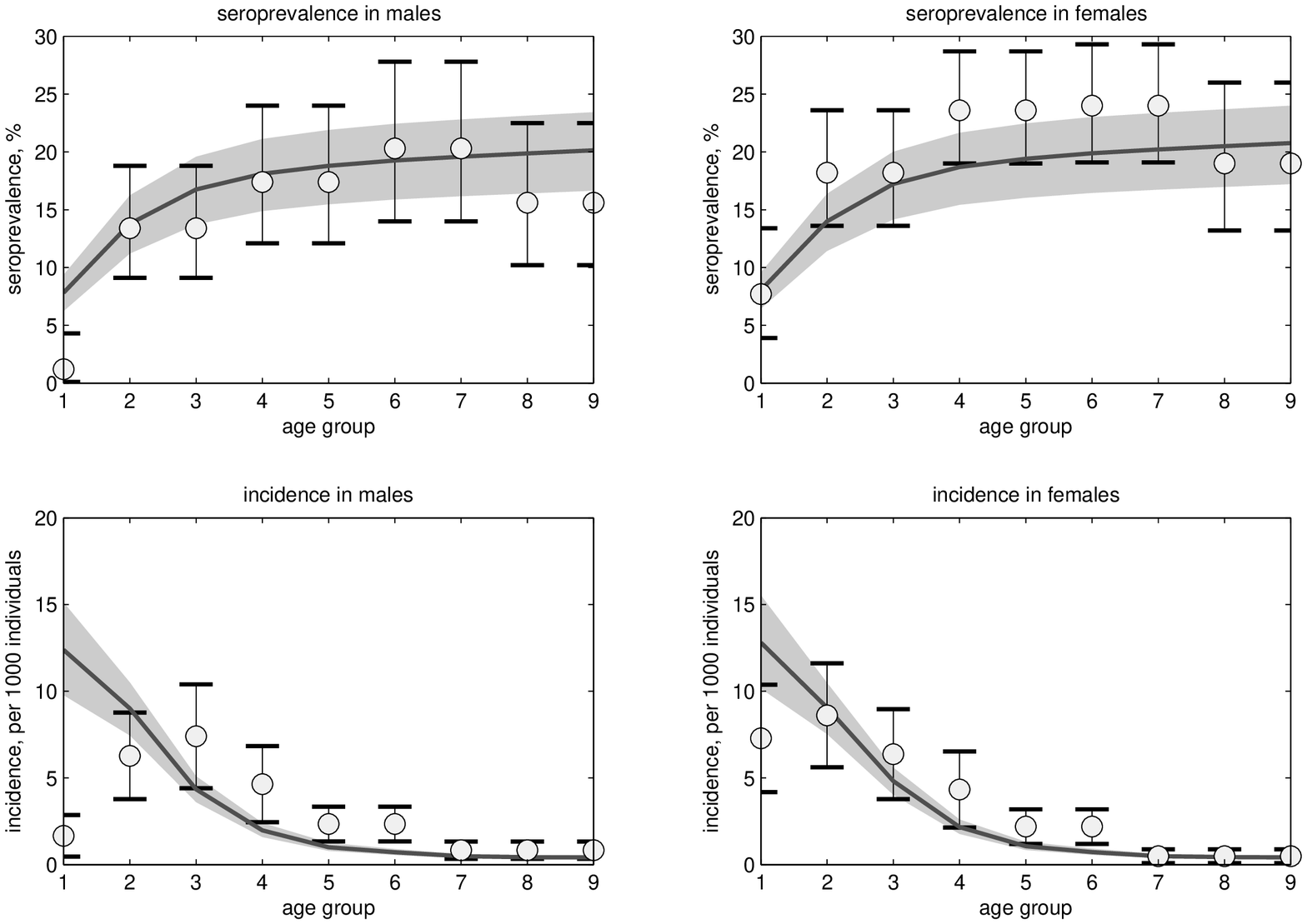}
\caption{Calibration for combined HPV-6 and -11: MMSE (solid line) and 95\% CI (shaded area).}
\label{fig:calibration_hpv611}
\end{center}
\end{figure}

 The first results we consider are for HPV-6 ( Fig. \ref{fig:calibration_hpv6}). The seroprevalence calibration is good for all age-groups except the youngest males and females in age-groups 4 and 5. At this point it is appropriate to note that all data recordings and observations we obtained for the males in age-group 1 can be questioned. We are inclined to think that the reported numbers do not represent the real-life situation since they suggest that both seroprevalence and incidence are  much lower among males in this age-group compared to the females of the same age. Considering the reported sexual partner acquisition rates (see Table \ref{tab:relative_rate_by_age_group}), which are the same for males and females in age-group 1, there is no apparent reason to believe this should be true and the reported numbers may simply be due to young males being reluctant to seek medical assistance. As for the females aged 30-39, their high seroprevalence level was not captured by the model with the prior distributions specified as in Table \ref{tab:used_priors}. Regarding incidence, we managed to obtain a reasonable fit for both females and males. The calibration for HPV-11 (Fig. \ref{fig:calibration_hpv11}) was accurate for seroprevalence (note that this is not the case for the youngest males and females though) and less accurate for incidence, especially when considering males. A similar result was observed for the combined HPV-6 and -11 data (Fig. \ref{fig:calibration_hpv611}).
 \newpage

\section{Impact of Vaccination via the Posterior Predictive Distribution}
In this section we demonstrate how to further develop the modelling framework provided for the calibration of the models developed for HPV-6 and -11 to incorporate a statistical analysis of the impact of vaccination. Having undertaken a calibration, via Adaptive MCMC forward projection methodology, presented in Section \ref{AdMCMCMethodology}, we obtain an empirical estimation of the posterior distribution for the model parameters presented in Table \ref{CoeffPrior} and the associated posterior estimates for the states at time $T$ when the last observation was taken prior to a vaccination treatment. This information can be represented by the empirical estimation of the marginal posterior distribution:
$$
p\left(\textbf{TR}, \textbf{WIP},\textbf{GWT}, \textbf{DAI}, \textbf{PSC}, \textbf{DI}, \Sigma, A_Y, EPSa, EPSr,\bm{X}_{g,a,s}(T)| \bm{O}_{g,a,s}(1),\ldots,\bm{O}_{g,a,s}(T)\right),
$$ 
using the $N$ samples from the Markov chain generated by the Adaptive MCMC Forward Projection algorithm. This empirical distribution can then be utilised to obtain the posterior predictive distribution, given by the resulting Monte Carlo approximation: 

\begin{multline}
\label{PostPred}
p(\bm{O}_{g,a,s}(T+1), \ldots,\bm{O}_{g,a,s}(T+\tau)|\bm{O}_{g,a,s}(1),\ldots,\bm{O}_{g,a,s}(T))= \\
\int p(\bm{O}_{g,a,s}(T+1),\ldots,\bm{O}_{g,a,s}(T+\tau)| \textbf{TR}, \textbf{WIP},\textbf{GWT}, \textbf{DAI}, \textbf{PSC}, \textbf{DI}, \Sigma, A_Y, EPSa, EPSr,\\
\bm{X}_{g,a,s}(T),\ldots,\bm{X}_{g,a,s}(T+\tau))\times p\left(
\textbf{TR}, \textbf{WIP},\textbf{GWT}, \textbf{DAI}, \textbf{PSC}, \textbf{DI}, \Sigma, A_Y, EPSa, EPSr,\right. \\
\left. \bm{X}_{g,a,s}(1),\ldots,\bm{X}_{g,a,s}(T)| \bm{O}_{g,a,s}(1),\ldots,\bm{O}_{g,a,s}(T)
\right)d\textbf{TR} \; d\textbf{WIP}\; d\textbf{GWT}\; d\textbf{DAI}\; d\textbf{PSC}\; d\textbf{DI}\;\\
 d\Sigma\; dA_Y\; dEPSa\; dEPSr= \frac{1}{J}\sum_{j=1}^J p
\left( 
\bm{O}_{g,a,s}(1),\ldots,\bm{O}_{g,a,s}(T)|\left\{\textbf{TR}, \textbf{WIP},\textbf{GWT}, \textbf{DAI}, \textbf{PSC}, \textbf{DI},\right.\right. \\
\left.\left. \Sigma, A_Y, EPSa, EPSr,\bm{X}_{g,a,s}(1),\ldots,\bm{X}_{g,a,s}(T)\right\}^j\right).
\end{multline}
Here the quantities 
\begin{align*}
p(\bm{O}_{g,a,s}(1),\ldots,\bm{O}_{g,a,s}(T)| \{\textbf{TR}, \textbf{WIP},\textbf{GWT}, \textbf{DAI}, \textbf{PSC}, \textbf{DI}, \Sigma, A_Y, EPSa, EPSr,\\
\bm{X}_{g,a,s}(1),\ldots,\bm{X}_{g,a,s}(T)\}^j)
\end{align*}
are obtained using the samples from the Adaptive MCMC Forward Projection calibration, then used to forward project the ODE solver to obtain estimates of $\bm{O}_{g,a,s}(T+1),\ldots,\bm{O}_{g,a,s}(T+\tau),$ conditional on sampled parameters and states at time $T$.

\begin{figure}[h!]
\begin{center}
\includegraphics[scale = 0.45]{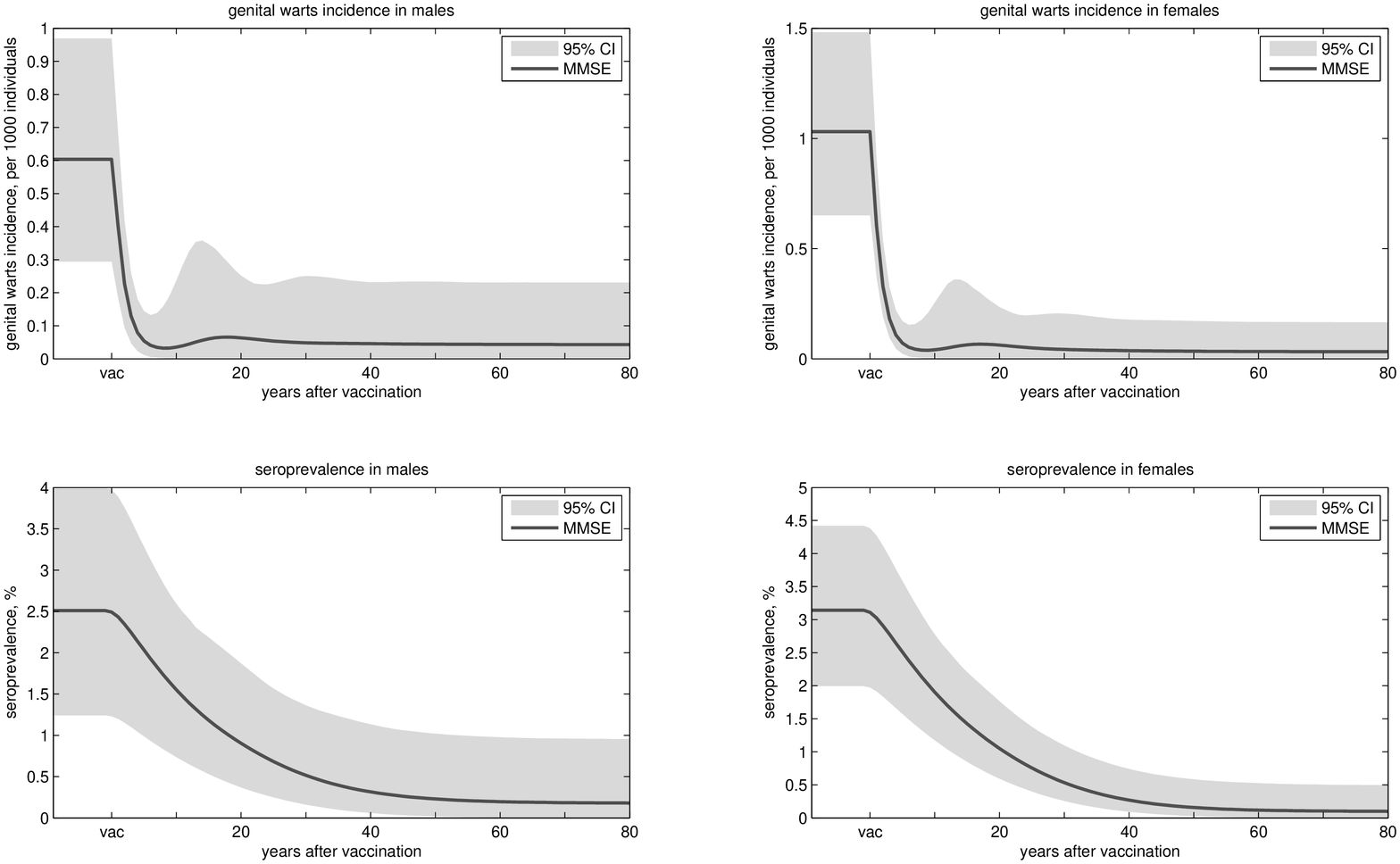}
\caption{Posterior predictive distribution for vaccine response: MMSE (solid line) and 95\% CI (shaded area).}
\label{fig:observations_after_vaccination}
\end{center}
\end{figure}

The results of this analysis are presented in Figure \ref{fig:observations_after_vaccination}. Here we present the posterior predictive distribution and predictive 95\% posterior confidence intervals for the incidence of HPV-6 and HPV-11 combined in the population post a vaccination. The vaccination scheme, commenced at the time $T+10$, assumes that 80\% of all 15-19 y.o. previously not vaccinated females (age-group 1) receive a single dose of vaccine each year. The effect of the vaccine is exclusively in reducing the force of infection on a vaccinated individual by 90\%, comparing to that on an individual belonging to the same age and sexual activity-group who has not been vaccinated. 

The results presented in this section have demonstrated how to incorporate the information that was obtained from the posterior calibration of the HPV-6 and HPV-11 models to observed data, to predict the vaccine response post calibration to a given equilibrium in the epidemic model within the population. They demonstrate the resultant new equilibrium level for incidence of HPV that would arise post vaccination, and the time taken to reach a new equilibrium level post-vaccination.

\section{Conclusion}
In this paper we developed a novel combination of a compartmental HPV-6/-11 ODE transmission model and a Bayesian statistical modelling framework. We investigated the possibility of calibrating the model to available seropositivity \cite{Newall2008}  and genital warts incidence \cite{Pirotta2009} data. We then demonstrated how to perform posterior predictive inference related to impact of vaccination on the modelled population. 

The estimation of the ODE model was achieved via adaptive Markov chain Monte Carlo based sampling methodology based around adaptive Metropolis coupled with a forward projection ODE solver to perform the joint challenge of sampling the static model parameters together with the estimated latent ODE states from transience to equilibrium. This was required in order to perform the evaluation of the rejection stage of the MCMC algorithm. We demonstrate the suitability of this sampling and estimation methodology and discuss the practical application of this estimation procedure. In the process we illustrate the capability of the proposed estimation procedure to perform calibration of these very high dimensional models for Australian data on observations of genital warts incidence and seroprevalence. We then develop a novel Bayesian perspective on studying the estimated vaccine response from such a model calibration, based around the posterior predictive distribution of the equilibrium states of the HPV epidemic model after vaccination.

In each case our aim was to ensure that the models developed were both parsimonious with respect to the number of parameters utilised to parametrize the ODEs describing the evolution of the disease epidemic for HPV-6 and -11 and sufficiently flexible to statistically calibrate the ODE compartmental model to the observed number of new genital warts diagnoses and seroprevalence per age-group. 

In addition, we maintain that the introduction of the approach we develop for modelling seroprevalence and its interpretation in the context of the poorly understood implications of seroconversion are in line with now popular views relating to HPV epidemic modelling. In particular, we assumed that only a person who recovered from either an asymptomatic HPV infection or genital warts, and is not susceptible, can be seropositive. 

An attempt to calibrate to the seroprevalence data was of interest because there were no such data available for Australia until they were presented in \cite{Newall2008}. Additionally, we were not aware of any published models calibrated to both seroprevalence and genital warts incidence data. Taking into account that all the data are necessarily prone to various errors we considered a calibration process successful if the simulated observation values were found within the reported 95\% confidence interval for the real data.

We have demonstrated that a reasonable calibration for the ODE epidemic model under consideration could be achieved only if some of the model parameters were allowed to take values beyond the ranges specified for them using currently available literature. We believe that this may be due to one or any  combination of the following reasons: an insufficient amount of HPV natural history detail incorporated in the model, poor quality of the available data and the uncertainty in interpretation of these data in the context of compartmental ODE models. Also, we have found that a notably better calibration can be observed if we change the assumptions we made about seropositivity in favour of a somewhat controversial idea that the seropositive status should persist regardless of an individual's disease state, and consider relatively short (up to 10 years) durations of immunity for both males and females. However, in this case the ODE system we had to deal with appeared to be stiff and the solver we used was significantly slower to complete solutions. An improvement to the quality of calibration can be ensured by the introduction of different durations of immunity for different age-groups. One concern with this extension to the model is the parsimony of the model, since the introduction of separate durations of immunity for all age-groups involves the necessity to deal with at least nine new parameters with non-informative prior distributions (or 18 new parameters if we want to improve the calibration further and assume different durations of immunity for males and females). Not only would one question in such settings the ability to accurately estimate such parameters in practice due to the observational data properties and the amount of observation data, but there would also be important statistical questions to be addressed relating to over parameterization of the model, over fitting of the model and parsimony. Since the intention of this work was to utilise parsimonious and interpretable models, we feel the model assumptions introduced were warranted to illustrate the properties of our estimation methodology. In addition, the extension of this methodology to working with additional parameterizations is trivial and easily incorporated into our estimation framework. Therefore we decided, based on the observed data under analysis, that it was prudent to maintain the parsimonious model representation developed.

On the other hand, we would like to emphasize that our proposed estimation methodology based around adaptive MCMC forward projection is automated and easily extended to facilitate any number of additional model parameters. This is in contrast to other approaches already proposed in the literature which may suffer from the curse of dimensionality under such parameter extensions. For example, had we utilised a basic Metropolis-Hastings sampler with a non-adaptive proposal mechanism, and not the adaptive Metropolis methodology we developed for this solution, then the tunning of the resulting transition kernel would be affected by such changes and would have to be performed again under each such model change. This would entail a significant computational burden in performing the model calibration under each change. In addition it may result in a serious practical hurdle to overcome in re-design of the proposal mechanism in the Metropolis-Hasting sampler in order to facilitate efficient mixing in such high dimensional sampling frameworks. 

We also note that our method is favorable relative other approaches which are not Markov chain based solutions. For example, Latin Hyper Cube sampling solutions which calibrate epidemic ODE models using grids over regions of the parameter space will suffer when the parameter space is significantly increased, see examples in HIV epidemic models such as \ref{blower1994sensitivity}. We note that under such approaches, an increase in the parameter space by an additional $m$ dimensions to represent the new model parameterization would require a grid be placed over an additional $m$ dimensional space, such significant increases in dimension may cause such approaches to suffer a curse of dimensionality attributed to the super-linear (exponential) increase in volume associated with adding the extra $m$ dimensions to the parameter space. For each such grid point, such techniques would then require a complete iteration of the ODE solver to stationarity, incurring significant additional computational cost for such model changes, even when sparse grid based techniques are used. Often incurring this additional computational cost for little benefit as regions of the parameter space of little significance to the most likely calibration points must be explored. This is in contrast to our adaptive MCMC based procedure which focus computational effort on regions of the parameter space proportionally with how probable, with respect to the posterior support, they are to contribute to suitable calibrations of the ode epidemic model.

In addition we mention that since LHC schemes do not straightforwardly facilitate a Bayesian sampling paradigm, and any probabilistic selection of grid points in the additional parameter space can be at best selected under such schemes based around prior information and perhaps heuristics on a goodness of fit criterion, then one readily realises the utility and flexibility of our proposed Bayesian framework and Adaptive MCMC Forward Projection methodology. This methodology is flexible in any dimension and under any model change, learning on-line the appropriate parameter subspaces to focus computational effort as dictated by a statistically consistent combination of both prior belief on rates of population movement between states and observed evidence. Therefore we are able to clearly quantify the probability weighted distribution of plausible calibration regions in the high dimensional parameter space using the samples we obtained from the posterior constructed. This is in contrast to previous literature which tries to get such information heuristically from procedures such as sensitivity analysis of the ode structure to different parameter regions, again incurring significant computational effort. Furthermore, at best such approaches can only produce what is termed the "variability or the prediction imprecision"  in the outcome variables that is due to the uncertainty in estimating the input values. It is important to note that such quantifications are not statistically based measures of prediction uncertainty, where as the predictive performance we develop under our Bayesian framework can be directly interpreted as the predictive distribution of the model, again providing advantages of our proposed Bayesian estimation approach in interpretation of the results.

\subsection{Acknowledgements}
We thank Dr David Philp for discussions and early work that motivated the work conducted for this paper. Mr Greg Londish is also thanked for work conducted in the early stages of this project. Funding for this project and support for IK were provided by an Australian Research Council Linkage Project grant (LP0883831). DR is supported by a National Health and Medical Research Council Program Grant (568971). The Kirby Institute is funded by the Australian Government Department of Health and Ageing and is affiliated with the Faculty of Medicine, University of New South Wales.

\section*{Appendix 1. Sexual mixing matrix}
Sexual mixing matrices are widely used in modelling of sexually transmitted diseases and their main purpose is to describe how certain characteristics of an individual define his or her sexual activity (i.e. the rate of sexual partner change). These characteristics typically are the age and/or sexual activity-group an individual belongs to.

\subsubsection*{Notations and available data }
From now on we will denote a gender by $g$, the gender opposite to $g$ by $g'$, a sexual activity (risk) group an by $s$ or $s'$ and $a$ or $a'$ will be an age-group. Also, when referring to an individual of gender $g$ who belongs to a sexual activity-group $s$ and age-group $a$ we will simply say "an individual from $g,s,a$" for brevity.

In the equations describing our model we have a force of infection term for every individual from $g,s,a$:
\begin{equation}
\label{def:force_of_infection_appendix}
\lambda_{g,s,a}=\beta_g\sum_{s',a'}\left\{c_{g,s,s',a,a'}\rho_{g,s,s',a,a'} \frac{I_{g',s',a'}}{ S_{g',s',a'}+ I_{g',s',a'}+ G_{g',s',a'}+ P_{g',s',a'}+ N_{g',s',a'}}\right\},
\end{equation}
where $\beta_g$ is the transmission probability per partnership, i.e. the probability that a susceptible person of gender $g$ will get infected from his or her infectious partner of gender $g'$; $c_{g,s,s',a,a'}$ is the mean per capita rate per year at which an individual from $g,s,a$ acquires new sexual partners from $g',s',a'$;  $\rho_{g,s,s',a,a'}$ is the conditional probability that an individual from $(g,s,a)$ forms a partnership with someone from  $g',s',a'$.
Products $c_{g,s,s',a,a'}\rho_{g,s,s',a,a'}$  are the rates of new sexual partner acquisition and they can be conveniently gathered in a matrix which we call a sexual mixing matrix.

Now we will start constructing a sexual mixing matrix for our model according to the approach by Garnett and Anderson \cite{Garnett1993} and its version used in \cite{Elbasha2008}. In the process of construction we should use the relevant data estimated from the results of Australian Study of Health and Relationships (ASHR). ASHR was a telephone survey of a random sample of about 20,000 people who were Australian residents aged from 16 to 59 y.o. Despite its limitations (see \cite{Smith2003} and  \cite{Regan2007} for discussion) this survey provided important information on sexual behaviour of Australians and its results are currently the most representative ones available. The data we need (as suggested in \cite{Garnett1993}, \cite{Garnett1993a} or \cite{Garnett1994}) are the following:
\begin{itemize}
\item relative sexual partner acquisition rates $r_a$ for each age-group $a$ (the same for males and females; see Table \ref{tab:relative_rate_by_age_group})
\item relative sexual partner acquisition rates $r_s$ for each sexual activity (risk) group $s$ (the same for males and females; Table \ref{tab:relative_rate_by_sex_group})
\item overall sexual partner acquisition rate $\bar{c}$ for the whole population (both males and females):
$$\bar{c}=0.437.$$
\end{itemize}
All these rates are per capita per year.
If for all $g,s,a$ we knew the rates at which an individual from $g,s,a$ acquires new sexual partners from the entire pool of individuals of gender $g'$ (denote them $c_{g,s,a}$), we could find the lowest of these rates,
$$
c_{min}:\,\, \forall g,s,a \quad c_{min}\le c_{g,s,a},
$$
and divide by it all $c_{g,s,a}$ denoting the results of division by $r_{g,s,a}$:
\begin{equation}
\label{eq:relative_rates_via_cmin}
r_{g,s,a}=\frac{c_{g,s,a}}{c_{min}}.
\end{equation}
These would be the relative sexual partner acquisition rates for $g,s,a$. We can specify them as follows
\begin{equation}
\label{def:relative_sex_partner_acquisition_rates}
r_{g,s,a}=r_a\times r_s \quad \forall g,s,a.
\end{equation}

We can now calculate $c_{min}$. Note that $\bar{c}N_g$ is the number of partners of gender $g'$ all individuals of gender $g$ acquire per year.
 This number must be equal to the total of sexual partners of gender $g'$ individuals of gender $g$ from each age and sex group acquire, so
$$\bar{c}N_g=\sum_{s,a}c_{g,s,a}N_{g,s,a}=c_{min}r_{g,1,1}N_{g,1,1}+c_{min}r_{g,1,2}N_{g,1,2}+\ldots=c_{min}\sum_{s,a}{r_{g,s,a}N_{g,s,a}}.$$
Hence,
\begin{equation}
\label{eq:cmin}
c_{min}=\frac{\bar{c}N_g}{\sum_{s,a}r_{g,s,a}N_{g,s,a}},
\end{equation}
also, with $c_{min}$ at hand we can easily calculate all $c_{g,s,a}$ (see (\ref{eq:relative_rates_via_cmin})).

\subsubsection*{Proportionate and assortative mixing}
Now we would like to specify $\rho_{g,s,s',a,a'}$, which are the conditional probabilities that an individual of gender $g$ from sexual activity-group $s$ and age-group $a$ gets a sexual partner of the opposite gender $g'$ from sexual activity-group $s'$ and age-group $a'$. There are three possibilities to consider:
\begin{enumerate}
	\item so-called "proportionate" sexual mixing by age-group or sexual activity-group; this means that $\rho_{g,s,s',a,a'}$ may be assumed equal to the proportion of  partnerships "generated" by individuals of gender $g'$  from age-group $a'$ (and/or sexual activity-group $s$):
\begin{equation}
\label{rho_proportional_age}
\rho_{g,s,s',a,a'}=\frac{\sum_{s'}c_{g',s',a'}N_{g',s',a'}}{\sum_{s',a'}c_{g',s',a'}N_{g',s',a'}}
\end{equation}
or
\begin{equation}
\label{rho_proportional_sex_group}
\rho_{g,s,s',a,a'}=\frac{\sum_{a'}c_{g',s',a'}N_{g',s',a'}}{\sum_{s',a'}c_{g',s',a'}N_{g',s',a'}}.
\end{equation}
In other words, an underlying assumption here is that more active members of gender $g'$ have higher chances to become a new sexual partner to someone of gender $g$. Note that in case of proportionate mixing by both age and sexual activity-group we should simply define $\rho_{g,s,s',a,a'}$ and a product of the right-hand sides of (\ref{rho_proportional_age}) and (\ref{rho_proportional_sex_group}).

\item "assortative" mixing (also known as "with-like" mixing); again, this can be by age or sexual activity-group or both, and if it's, for example,  by age-group, we define
\begin{equation}
\label{rho_assortative_age}
\rho_{g,s,s',a,a'}=\delta_{aa'}.
\end{equation}
That is, the probability of establishing a partnership is 1 if a potential partner is of the same age and 0 otherwise.
\item "disassortative" mixing ("with-unlike") suggests that a partnership can be formed only with someone from a different age (or sexual activity) group.
\end{enumerate}

We would like to be able to adjust the sort of mixing in our model depending on our needs. Therefore, we specify the probabilities $\rho_{g,s,s',a,a'}$ as follows
\begin{align}
\rho_{g,s,s',a,a'}=\left(
(1-EPSa)\delta_{aa'}+EPSa\frac{\sum_{s'}c_{g',s',a'}N_{g',s',a'}}{\sum_{s',a'}c_{g',s',a'}N_{g',s',a'}}
\right)
\times \nonumber\\
\left(
(1-EPSr)\delta_{ss'}+EPSr\frac{\sum_{a'}c_{g',s',a'}N_{g',s',a'}}{\sum_{s',a'}c_{g',s',a'}N_{g',s',a'}}.
\right)
\end{align}

Parameters $EPSa$ and $EPSr$ help us vary the extent of assortativeness by age and sexual activity-group: if $EPSa=0$ mixing is fully assortative by age-group; if  $EPSa=1$
it is fully proportionate by age-group. In a similar fashion we can vary $EPSr$ and control assortativeness by sexual activity-group.

\subsubsection*{Age related adjustments}
Here we introduce some adjustments to emphasise the effect of a steady popularity of the partnerships between older males and younger females. Let for now $g$ denote males and $g'$ females. We reduce the probabilities that males in the age-group 3 and higher have female partners of the same age:
\begin{equation}
\label{sameage_males}
\rho_{g,s,s',a,a'}\rightarrow \rho_{g,s,s',a,a'}(1-\Gamma)\quad \mbox{if }\left\{\begin{array}{l}a= a'\\ a\ge 3\end{array}\right.
\end{equation}
Similarly, we reduce the probabilities that females in the age-groups 1 to 5 form a partnership with males of the same age:
\begin{equation}
\label{sameage_females}
\rho_{g',s,s',a,a'}\rightarrow \rho_{g',s,s',a,a'}(1-\Gamma)\quad \mbox{if }\left\{\begin{array}{l}a= a'\\ a\le 5\end{array}\right.
\end{equation}
To compensate for the effect of these adjustments we increase the probabilities for males to have a female partner one age-group younger but belonging to the age-groups not higher than 5,
\begin{equation}
\label{males_like_younger_females}
\rho_{g,s,s',a,a'}\rightarrow \rho_{g,s,s',a,a'}+\Gamma\rho_{g,s,s',a,a}\quad \mbox{if }\left\{\begin{array}{l}a=a'+2\\ a'\le 5\end{array}\right.
\end{equation}
and also we increase the probabilities that female have one age-group older males from the age-group 3 or higher as a partner:
\begin{equation}
\label{females_like_older_males}
\rho_{g',s,s',a,a'}\rightarrow \rho_{g',s,s',a,a'}+\Gamma\rho_{g',s,s',a,a}\quad \mbox{if }\left\{\begin{array}{l}a=a'-2\\ a'\ge 3\end{array}\right.
\end{equation}

So far we have used the rates $c_{g,s,a}$ which describe new sexual partner acquisitions by individuals from $g,s,a$ from the whole population of gender $g'$. So to find out the rates of acquisition of new sexual partners from
a certain age and sexual activity-group ($a'$ and $s'$) we should multiply  $c_{g,s,a}$ by $\rho_{g,s,s',a,a'}$. However, now we would like to make the rates $c_{g,s,a}$ group-specific in terms of the groups the sexual partners are selected from. This will be achieved via balancing supply and demand of sexual partners.
\subsubsection*{Balancing supply and demand}
We want the following to hold for all $g,s,a$ and $g',s',a'$:
\begin{equation}
\label{supply_and_demand}
c_{g,s,a}\rho_{g,s,s',a,a'}N_{g,s,a}=c_{g',s',a'}\rho_{g',s',s,a',a}N_{g',s',a'}.
\end{equation}
Here $c_{g,s,a}\rho_{g,s,s',a,a'}$ is the rate of acquisition of new sexual partners by individuals who belong to $g,s,a$ from $g',s',a'$, and  $c_{g,s,a}\rho_{g,s,s',a,a'}N_{g,s,a}$ is the total number of new partners acquired by $g,s,a$ from $g',s',a'$. So the equation simply states that  the total number of new partners acquired by $g,s,a$ from $g',s',a'$ must be equal to  the total number of new partners acquired by $g',s',a'$ from $g,s,a$.

Let
$$
c_{g,s,a}\rho_{g,s,s',a,a'}N_{g,s,a}\ne c_{g',s',a'}\rho_{g',s',s,a',a}N_{g',s',a'}.
$$
We want to find such a multiplier $B$ that
$$
B^{\theta_1}c_{g,s,a}\rho_{g,s,s',a,a'}N_{g,s,a}=B^{\theta_2} c_{g',s',a'}\rho_{g',s',s,a',a}N_{g',s',a'}.
$$

Then
$$
B^{\theta_1-\theta_2}=\frac{c_{g',s',a'}\rho_{g',s',s,a',a}N_{g',s',a'}}{c_{g,s,a}\rho_{g,s,s',a,a'}N_{g,s,a}}.
$$
To keep things simple, let $\theta_1-\theta_2=1.$  Note that $B$ serves as a degree of imbalance: the balance is established if $B=1$. So, to ensure that (\ref{supply_and_demand}) is true it is enough to
introduce the group-specific rates  $c_{g,s,s',a,a'}$ to be used instead of $c_{g,s,a}$:
\begin{equation}
\label{balanced_rates}
c_{g,s,s',a,a'}=c_{g,s,a}B^{\theta_1}, \quad c_{g',s',s,a',a}=c_{g',s',a'}B^{\theta_1-1}
\end{equation}
We limit the range of value of parameter $\theta_1$ to $[0,1]$. Suppose the supply and demand are not balanced and $\theta_1=0$. Then as follows from (\ref{balanced_rates}), $c_{g,s,s',a,a'}=c_{g,s,a}$ (the rates for gender $g$ do not get adjusted), but $c_{g',s',s,a',a}=c_{g',s',a'}B^{-1}$ (the rates for gender $g'$ are adjusted). If $\theta_1=1$ it is the other way around. Consequently, $\theta_1$ indicates to what extent individuals of each gender adjusts their sexual partner acquisition rates (i.e. we may say, sexual behaviour) in case the available supply of sexual partners does not meet demand.

At this stage our sexual mixing matrix is fully formed.

\bibliography{statmed.paper.references}

\begin{thebibliography}{10}

\bibitem{Anderson1992}
Roy~M. Anderson and Robert~M. May.
\newblock {\em Infectious Diseases of Humans: Dynamics and Control}.
\newblock Oxford University Press, 1992.

\bibitem{Andrieu2006}
C.~Andrieu and .Y.~F. Atchade.
\newblock On the efficiency of adaptive {MCMC} algorithms.
\newblock In {\em Proceedings of the 1st {I}nternational {C}onference on
  Performance {E}valuation {M}ethodolgies and {T}ools}, page~43. ACM, 2006.

\bibitem{Andrieu2008}
C.~Andrieu and J.~Thoms.
\newblock A tutorial on adaptive {MCMC}.
\newblock {\em Statistics and Computing}, 18(4):343--373, 2008.

\bibitem{Arima2010}
Yuzo Arima, Rachel~L. Winer, Qinghua Feng, James~P. Hughes, Shu-Kuang Lee,
  Michael~E. Stern, Sandra~F. O’Reilly, , and Laura~A. Koutsky.
\newblock Development of genital warts after incident detection of human
  papillomavirus infection in young men.
\newblock {\em The Journal of Infectious Diseases}, 202:1181--1184, 2010.

\bibitem{Atchade2009}
Y.~Atchad{\'e}, G.~Fort, E.~Moulines, and P.~Priouret.
\newblock Adaptive {M}arkov chain {M}onte {C}arlo: Theory and methods.
\newblock 2009.

\bibitem{Atchade2005}
Y.F. Atchad{\'e} and J.S. Rosenthal.
\newblock On adaptive {M}arkov chain {M}onte {C}arlo algorithms.
\newblock {\em Bernoulli}, 11(5):815--828, 2005.

\bibitem{Barnabas2006}
Ruanne~V. Barnabas, Paivi Laukkanen, Pentti Koskela, Osmo Kontula, Matti
  Lehtinen, and Geoff~P. Garnett.
\newblock Epidemiology of {HPV} 16 and cervical cancer in {F}inland and the
  potential impact of vaccination: Mathematical modelling analyses.
\newblock {\em PLoS Medicine}, 3:624--632, 2006.

\bibitem{Beutels2010}
P.~Beutels and M.~Jit.
\newblock A brief history of economic evaluation for human papillomavirus
  vaccination policy.
\newblock {\em Sexual health}, 7(3):352–358, 2010.

\bibitem{Brisson2003}
M.~Brisson and W.~J. Edmunds.
\newblock Economic evaluation of vaccination programs: The impact of
  herd-immunity.
\newblock {\em Med. Decis. Mak.}, 23(1):76--82, 2003.

\bibitem{brooks1998markov}
S.~Brooks.
\newblock {M}arkov chain {M}onte {C}arlo method and its application.
\newblock {\em Journal of the Royal Statistical Society: Series D (The
  Statistician)}, 47(1):69--100, 1998.

\bibitem{Burchell2006}
Ann~N. Burchell, Harriet Richardson, Salaheddin~M. Mahmud, Helen Trottier,
  Pierre~P. Tellier, James Hanley, Francois Coutlee, and Eduardo~L. Franco.
\newblock Modeling the sexual transmissibility of human papillomavirus
  infection using stochastic computer simulation and empirical data from a
  cohort study of young women in {M}ontreal, {C}anada.
\newblock {\em American Journal of Epidemiology}, 163(6):534--543, 2006.

\bibitem{Burchell2006a}
Ann~N. Burchell, Rachel~L. Winer, Silvia de~Sanjose, and Eduardo~L. Franco.
\newblock Epidemiology and transmission dynamics of genital {HPV} infection.
\newblock {\em Vaccine}, 24, Suppl 3:S52--S61, 2006.

\bibitem{Choi2010}
Y.~H. Choi, M.~Jit, N.~Gay, A.~Cox, G.~P. Garnett, and W.~J. Edmunds.
\newblock Transmission dynamic modelling of the impact of human papillomavirus
  vaccination in the {U}nited {K}ingdom.
\newblock {\em Vaccine}, 28(24):4091--4102, 2010.

\bibitem{DitlevsenDegaetano2005mixed}
S.~Ditlevsen and A.~De~Gaetano.
\newblock {Mixed effects in stochastic differential equation models}.
\newblock {\em REVSTAT-Statistical Journal}, 3(2):137--153, 2005.

\bibitem{DitlevsenDegaetano2005rat}
S.~Ditlevsen and A.~De~Gaetano.
\newblock Stochastic vs. deterministic uptake of dodecanedioic acid by isolated
  rat livers.
\newblock {\em Bulletin of mathematical biology}, 67(3):547--561, 2005.

\bibitem{Edmunds1999}
W.~J. Edmunds, G.~F. Medley, and D.~J. Nokes.
\newblock Evaluating the cost-effectiveness of vaccination programmes: A
  dynamic perspective.
\newblock {\em Stat. Med.}, 18(23):3263--3282, 1999.

\bibitem{Elbasha2007}
Elamin~H. Elbasha, Erik~J. Dasbach, and Ralph~P. Insinga.
\newblock Model for assessing human papillomavirus vaccination strategies.
\newblock {\em Emerging Infectious Diseases}, 13(1):28--41, 2007.

\bibitem{Elbasha2008}
Elamin~H. Elbasha, Erik~J. Dasbach, and Ralph~P. Insinga.
\newblock A multi-type {HPV} transmission model.
\newblock {\em Bulletin of Mathematical Biology}, 70:2126--2176, 2008.

\bibitem{Garland2010}
S.~M. Garland, J.~M. Brotherton, C.~K. Fairley, D.~M. Gertig, and M.~Saville.
\newblock Advancements in the control of genital human papillomavirus
  infections and related diseases: highlighting {A}ustralia's role.
\newblock {\em Sex Health}, 7(3):227--229, 2010.

\bibitem{Garnett2005}
G.~P. Garnett.
\newblock Role of herd immunity in determining the effect of vaccines against
  sexually transmitted disease.
\newblock {\em J. Infect. Dis.}, 191, Suppl 1:S97--S106, 2005.

\bibitem{Garnett1993a}
Geoff~P. Garnett and Roy~M. Anderson.
\newblock Contact tracing and the estimation of sexual mixing patterns: the
  epidemiology of gonococcal infections.
\newblock {\em Sexually Transmitted Diseases}, 4:181--191, 1993.

\bibitem{Garnett1996}
Geoff~P. Garnett and Roy~M. Anderson.
\newblock Sexually transmitted diseases and sexual behavior: insights from
  mathematical models.
\newblock {\em Journal of Infectious Diseases}, 174:S150--S161, 1996.

\bibitem{Garnett1993}
Geoffrey~P. Garnett and Roy~M. Anderson.
\newblock Factors controlling the spread of {HIV} in heterosexual communities
  in developing countries: Patterns of mixing between different age and sexual
  activity classes.
\newblock {\em Philosophical Transactions of the Royal Society B: Biological
  Sciences}, 342:137--159, 1993.

\bibitem{Garnett1994}
Geoffrey~P. Garnett and Roy~M. Anderson.
\newblock Balancing sexual partnerships in an age and activity stratified model
  of {HIV} transmission in heterosexual populations.
\newblock {\em IMA Journal of Mathematics Applied in Medicine \& Biology},
  11:161--192, 1994.

\bibitem{Garnett1996a}
Geoffrey~P. Garnett, James~P. Hughes, Roy~M. Anderson, Bradley~P. Stoner,
  Sevgi~O. Aral, William~L. Whittington, Hunter~H. Handsfield, and King.~K.
  Holmes.
\newblock Sexual mixing patterns of patients attending sexually transmitted
  diseases clinics.
\newblock {\em Sexually Transmitted Diseases}, 23(3):248--257, 1996.

\bibitem{Ghani1997}
A.~C. Ghani, J.~Swinton, and G.~P. Garnett.
\newblock The role of sexual partnership networks in the epidemiology of
  gonorrhea.
\newblock {\em Sexually Transmitted Diseases}, 24(1):45--56, 1997.

\bibitem{Ghani1998}
A.C. Ghani and G.P. Garnett.
\newblock Measuring sexual partner networks for transmission of sexually
  transmitted diseases.
\newblock {\em Journal of the Royal Statistical Society: Series A (Statistics
  in Society)}, 161:227--238, 1998.

\bibitem{gilks1996markov}
W.R. Gilks, S.~Richardson, and D.J. Spiegelhalter.
\newblock {M}arkov chain {M}onte {C}arlo in practice.
\newblock 1996.

\bibitem{Giuliano2008}
Anna~R. Giuliano, Beibei Lu, Carrie~M. Nielson, Roberto Flores, Mary~R.
  Papenfuss, Ji-Hyun Lee, Martha Abrahamsen, and Robin~B. Harris.
\newblock Age-specific prevalence, incidence, and duration of human
  papillomavirus infections in a cohort of 290 {US} men.
\newblock {\em The Journal of Infectious Diseases}, 198:827--835, 2008.

\bibitem{Haario2001}
H.~Haario, E.~Saksman, and J.~Tamminen.
\newblock An adaptive {M}etropolis algorithm.
\newblock {\em Bernoulli}, 7:223--242, 2001.

\bibitem{Hoare2008}
A.~Hoare, D.~G. Regan, and D.~P. Wilson.
\newblock Sampling and sensitivity analyses tools ({SaSAT}) for computational
  modelling.
\newblock {\em Theoretical Biology and Medical Modelling}, 5:1--18, 2008.

\bibitem{Hughes2002}
J.~P. Hughes, G.~P. Garnett, and L.~Koutsky.
\newblock The theoretical population-level impact of a prophylactic human
  papilloma virus vaccine.
\newblock {\em Epidemiology}, 13(6):631--639, 2002.

\bibitem{Insinga2003}
Ralph~P. Insinga, Erik~J. Dasbach, and Evan~R. Myers.
\newblock The health and economic burden of genital warts in a set of private
  health plans in the {U}nited {S}tates.
\newblock {\em Clinical Infectious Diseases}, 36:1397--1403, 2003.

\bibitem{Jit2008}
Mark Jit, Yoon~Hong Choi, and W.~John Edmunds.
\newblock Economic evaluation of human papillomavirus vaccination in the
  {U}nited {K}ingdom.
\newblock {\em BMJ}, 337:a769, 2008.

\bibitem{Keeling2007}
M.J. Keeling and Pejman Rohani.
\newblock {\em Modeling Infectious Diseases in Humans and Animals}.
\newblock Princeton University Press, 2007.

\bibitem{Kim2008}
Jane~J. Kim and Sue~J. Goldie.
\newblock Health and economic implications of {HPV} vaccination in the {U}nited
  {S}tates.
\newblock {\em New England Journal of Medicine}, 359(8):821--832, 2008.

\bibitem{Kim2007}
J.J. Kim, B.~Andres-Beck, and S.J. Goldie.
\newblock The value of including boys in an {HPV} vaccination programme: a
  cost-effectiveness analysis in a low-resource setting.
\newblock {\em British Journal of Cancer}, 97:1322--1328, 2007.

\bibitem{Kulasingam2007}
Shalini Kulasingam, Luke Connelly, Elizabeth Conway, Jane~S. Hocking, Evan
  Myers, David~G. Regan, David Roder, Jayne Ross, and Gerard Wain.
\newblock A cost-effectiveness analysis of adding a human papillomavirus
  vaccine to the {A}ustralian national cervical screening program.
\newblock {\em Sexual health}, 4:165--175, 2007.

\bibitem{Lacey2006}
C.~J. Lacey, C.~M. Lowndes, and K.~V. Shah.
\newblock Burden and management of non-cancerous {HPV}-related conditions:
  {HPV}-6/11 disease.
\newblock {\em Vaccine}, 24, Suppl 3:S35--S41, 2006.

\bibitem{Marra2009}
F.~Marra, K.~Cloutier, B.~Oteng, C.~Marra, and G.~Ogilvie.
\newblock Effectiveness and cost effectiveness of human papillomavirus vaccine:
  a systematic review.
\newblock {\em Pharmacoeconomics}, 27(2):127--47, 2009.

\bibitem{Munoz2006}
N.~Munoz, X.~Castellsague, A.~B. de~Gonzalez, and L.~Gissmann.
\newblock {HPV} in the etiology of human cancer.
\newblock {\em Vaccine}, 24, Suppl 3:S1--S10, 2006.

\bibitem{Newall2008}
Anthony~T. Newall, Julia M.~L. Brotherton, Helen~E. Quinn, Peter~B. McIntyre,
  Josephine Backhouse, Lynn Gilbert, Mark~T. Esser, Joanne Erick, Janine Bryan,
  Neil Formica, and C.~Raina MacIntyre.
\newblock Population seroprevalence of human papillomavirus types 6, 11, 16,
  and 18 in men, women, and children in {A}ustralia.
\newblock {\em Clinical Infectious Diseases}, 46:1647--1655, 2008.

\bibitem{Pirotta2009}
Marie Pirotta, Alicia~Noemi Stein, E.~Lynne Conway, Christopher Harrison,
  Helena Britt, and Suzanne Garland.
\newblock Genital warts incidence and health care resource utilisation in
  {A}ustralia.
\newblock {\em Sexually Transmitted Infections}, 86:181--186, 2009.

\bibitem{Pomfret2011}
T.~C. Pomfret, J.~M. Gagnon~Jr, and A.~T. Gilchrist.
\newblock Quadrivalent human papillomavirus (hpv) vaccine: A review of safety,
  efficacy, and pharmacoeconomics.
\newblock {\em Journal of Clinical Pharmacy and Therapeutics}, 36(1):1--9--,
  2011.

\bibitem{Regan2007}
David~G. Regan, David~J. Philp, Jane~S. Hocking, and Matthew~G Law.
\newblock Modelling the population-level impact of vaccination on the
  transmission of human papillomavirus type 16 in {A}ustralia.
\newblock {\em Sex. Health}, 4:147--163, 2007.

\bibitem{Regan2010}
David~G. Regan, David~J. Philp, and Edward~K. Waters.
\newblock Unresolved questions concerning human papillomavirus infection and
  transmission: a modelling perspective.
\newblock {\em Sexual health}, 7:368--375, 2010.

\bibitem{Regan2008}
David~G. Regan and David~P. Wilson.
\newblock Modelling sexually transmitted infections: less is usually more for
  informing public health policy.
\newblock {\em Transactions of the Royal Society of Tropical Medicine and
  Hygiene}, 102(207-208):207--8, 2008.

\bibitem{Roberts2001}
Gareth~O. Roberts and Geoffrey~S. Rosenthal.
\newblock Optimal scaling for various {M}etropolis-{H}astings algorithms.
\newblock {\em Statistical Science}, 16(4):351--367, 2001.

\bibitem{Roberts2009}
Gareth~O. Roberts and Jeffrey~S. Rosenthal.
\newblock Examples of adaptive {MCMC}.
\newblock {\em Journal of Computational and Graphical Statistics}, 18:349--367,
  2009.

\bibitem{roberts2001optimal}
G.O. Roberts and J.S. Rosenthal.
\newblock Optimal scaling for various {M}etropolis-{H}astings algorithms.
\newblock {\em Statistical Science}, 16(4):351--367, 2001.

\bibitem{Saltelli2000}
A.~Saltelli, S.~Tarantola, and F.~Campolongo.
\newblock Sensitivity analysis as an ingredient of modeling.
\newblock {\em Statistical Science}, 15(4):377--395, 2000.

\bibitem{Sanders2003}
G.~D. Sanders and A.~V. Taira.
\newblock Cost effectiveness of a potential vaccine for human papillomavirus.
\newblock {\em Emerg. Infect. Dis}, 9(1):37--48, 2003.

\bibitem{Smith2003}
Anthony~M.A. Smith, Chris~E. Rissel, Juliet Richters, Andrew~E. Grulich, and
  Richard~O. de~Visser.
\newblock Sex in {A}ustralia: The rationale and methods of the {A}ustralian
  {S}tudy of {H}ealth and {R}elationships.
\newblock {\em Australian and New Zealand Journal of Public Health},
  27:106--117, 2003.

\bibitem{Stanley2010}
M.~Stanley.
\newblock Prophylactic human papillomavirus vaccines: will they do their job?
\newblock {\em Journal of Internal Medicine}, 267(3):251--259, 2010.

\bibitem{Trottier2009}
H.~Trottier and A.~N. Burchell.
\newblock Epidemiology of mucosal human papillomavirus infection and associated
  diseases.
\newblock {\em Public Health Genomics}, 12(5-6):291--307, 2009.

\bibitem{Trottier2008}
Helen Trottier, Salaheddin Mahmud, José {C}arlos M.~Prado, Joao~S. Sobrinho,
  Maria~C. Costa, Thomas~E. Rohan, Luisa~L. Villa, and Eduardo~L. Franco.
\newblock Type-specific duration of human papillomavirus infection:
  Implications for human papillomavirus screening and vaccination.
\newblock {\em The Journal of Infectious Diseases}, 197:1436--1447, 2008.

\bibitem{VandeVelde2007}
Nicolas Van~de Velde, Marc Brisson, and Marie-Claude Boily.
\newblock Modeling human papillomavirus vaccine effectiveness: quantifying the
  impact of parameter uncertainty.
\newblock {\em American Journal of Epidemiology}, 165(7):762--775, 2007.

\bibitem{Veldhuijzen2010}
N.~J. Veldhuijzen, P.~J. Snijders, P.~Reiss, C.~J. Meijer, and J.~H. van~de
  Wijgert.
\newblock Factors affecting transmission of mucosal human papillomavirus.
\newblock {\em Lancet Infectious Diseases}, 10(12):862--874, 2010.

\bibitem{Vynnycky2010}
Emilia Vynnycky and Richard~G. White.
\newblock {\em An Introduction to Infectious Disease Modelling}.
\newblock Oxford University Press, 2010.

\bibitem{Winer2005}
Rachel~L. Winer, Nancy~B. Kiviat, James~P. Hughes, Diane~E. Adam, Shu-Kuang
  Lee, Jane~M. Kuypers, and Laura~A. Koutsky.
\newblock Development and duration of human papillomavirus lesions, after
  initial infection.
\newblock {\em The Journal of Infectious Diseases}, 191:731--738, 2005.

\end{thebibliography}
\bibliographystyle{plain}

\begin{table}[h]
\centering
\begin{tabular}{ll}
\hline\hline
\textbf{Disease State} & \textbf{Description} \\
\hline\hline
$S$ & `\textsl{Susceptible}': an individual who is at risk of getting infected;\\
 & \\
$I$ & `\textsl{Infected}': an individual with type 6 or 11 HPV;\\
 & \\
$G$ & `\textsl{Genital warts}': an individual who developed genital warts and is being treated; \\

 & \\
$P$ & `\textsl{Seropositive}': an individual who is recovered and seropositive; \\
 & \\
$N$ & `\textsl{Seronegative}': an individual who is recovered and seronegative.\\
\hline\hline
\end{tabular}
\caption{Compartments defined in the HPV-6/-11 transmission model according to disease states within a population.}
\label{tab:compartments}
\end{table}

\begin{table}
\centering
\begin{tabular}{ll}
\hline\hline
\textbf{Coefficient} & \textbf{Coefficient Interpretation} \\
\hline\hline
$\lambda_{g,s,a}(t)$ & Force (risk) of infection - a per capita rate of \\
                  &   becoming infected per unit time (a year in our model); \\
 & \\
 $r_g$& Recovery rate for a patient who is being treated; approximately equal to \\
 & 1/(average duration of treatment for genital warts); \\
 & \\
$\rho_g$ & Recovery rate for an asymptomatic individual; approximately equal to\\
 & 1/(average duration of an untreated HPV infection);\\
 & \\
$\nu_g$ & Probability that an individual becomes seropositive;\\
 & \\
$\zeta_g$ & Rate at which an individual loses immunity; approximately equal\\
 & to 1/(average duration of immunity);\\
 & \\
$\gamma_g$ & Genital warts development rate; approximately equal to\\
             &1/(average time period between becoming infected and detection of genital warts). \\
\hline\hline
\end{tabular}
\caption{Interpretations of the coefficients in the HPV-6/-11 model equations.}
\label{tab:parameter_interpretation}
\end{table}

\begin{table}
\tabcolsep=0.11cm
\begin{tabular}{lll}
\hline \hline
\multicolumn{1}{c}{\textbf{Parameter}} & \multicolumn{1}{c}{\textbf{Interpretation}} & \multicolumn{1}{c}{\textbf{Prior}} \\ \hline \hline
\multicolumn{3}{c}{\textbf{Non-linear ODE Model Parameter Priors}}\\ \hline
Transmission probability (males) & TRm & $U[Ba,Bb]$\\
Transmission probability (females) & TRf & $U[Ba,Bb]$\\
Average genital warts incubation period  (males)  & WIPm & $Ga(k_{WIPm},\theta_{WIPm})$\\
Average genital warts incubation period  (females)& WIPf & $Ga(k_{WIPf},\theta_{WIPf})$\\
Average duration of genital warts treatment (males) & DWTm & $Ga(k_{DWTm},\theta_{DWTm})$\\
Average duration of genital warts treatment (females) & DWTf & $Ga(k_{DWTf},\theta_{DWTf})$\\
Average duration of asymptomatic HPV infection (males) & DAIm & $Ga(k_{DAIm},\theta_{DAIm})$\\
Average duration of asymptomatic HPV infection (females)  & DAIf & $Ga(k_{DAIf},\theta_{DAIf})$\\
Average duration of immunity (males) & DIm & $U(k_{DIm},\theta_{DIm})$\\
Average duration of immunity (females) & DIf & $U(k_{DIf},\theta_{DIf})$\\
Probability of seroconversion (males)& PSCm & $Be(\alpha_{PSCm},\beta_{PSCm})$\\
Probability of seroconversion (females)& PSCf & $Be(\alpha_{PSCf},\beta_{PSCf})$\\
\hline
\multicolumn{3}{c}{\textbf{Observation Error Parameters Priors}}\\ \hline
Diagonal element of the observation error covariance matrix & & \\
for incidence $\Sigma=\mbox{diag}(\sigma)$ & $\sigma$ & $invGa(k_{\sigma},\theta_{\sigma})$ \\
Observation error scale for seroprevalence & $A_Y$ & $Ga(k_{A_Y},\theta_{A_Y})$\\\hline
\multicolumn{3}{c}{\textbf{Mixing Matrix Parameter Priors}}\\ \hline
Degree of assortativeness by age-group & EPSa & $Be(\alpha_{\epsilon_a},\beta_{\epsilon_a})$\\
Degree of assortativeness by risk-group & EPSr & $Be(\alpha_{\epsilon_r},\beta_{\epsilon_r})$\\
\hline
\end{tabular}
\caption{\label{CoeffPrior} Prior specification for non-linear-ODE models for HPV epidemic. Note that the non-linear ODE model parameter priors can all be assumed non-informative.}
\end{table}
\begin{table}
\begin{tabular}{lccc|ccc}
 & \multicolumn{3}{c}{\textbf{males}} &\multicolumn{3}{c}{\textbf{females}}\\
\hline\hline
\textbf{duration (in years) of}& \textbf{value} & \textbf{95\%CI} & \textbf{source}   & \textbf{value} & \textbf{95\%CI} & \textbf{source}   \\
\hline\hline
Incubation period, median & 0.916 & 0-1.341 & \cite{Arima2010} & 0.241 & 0-0.475 &\cite{Winer2005}\\
Treatment, mean & 0.281 & 0.213-0.349 & \cite{Insinga2003} &0.232 & 0.185-0.280 & \cite{Insinga2003}\\
Treatment, median &  N/A & N/A & N/A & 0.491 & 0.325-0.666 & \cite{Insinga2003}\\
Untreated infection, median &  0.45 & 0.425-0.475 & \cite{Giuliano2008} & 0.5 & 0.475-0.575 & \cite{Trottier2008}\\
Untreated infection, mean &  N/A & N/A& N/A & 0.7916 & 0.575-1.0 & \cite{Trottier2008}\\
\hline
\end{tabular}
\caption{Data from medical literature which aid in specification of prior hyper-parameter values for the HPV-6/-11 model.}
\label{tab:real_data}
\end{table}
\begin{table}[h]
\centering
\begin{tabular}{ccccc}
\hline\hline
                       &        & $WIP$ &      $DAI$&           $DWT$ \\
\hline\hline
\multirow{2}{*}{6}   & males  &$U(0.9,1.3)$&$U(2.2,3.6)$&$Ga(92.00,0.003)$\\
                     & females&$U(0.6,0.9)$&$U(2.2,3.6)$&$Be(69.00,231.00)$ \\
\hline\hline
\multirow{2}{*}{11}  & males  &$U(0.9,1.3)$&$U(2.0,3.6)$&$Ga(92.00,0.003)$\\
                     & females&$U(0.6,0.9)$&$U(2.0,3.6)$&$Be(69.00,231.00)$ \\
\hline\hline
\multirow{2}{*}{6/11}& males  &$U(1.0,2.0)$&$U(3.8,4.8)$&$Ga(92.00,0.003)$\\
                     & females&$U(1.0,2.0)$&$U(3.8,4.8)$&$Be(69.00,231.00)$ \\
\hline\hline
\end{tabular}
\caption{Selected prior distributions actually used in our simulations. While $DWT$ is in agreement with the real data, $WIP$ and $DAI$ were allowed to take values beyond their reported ranges (see Table \ref{tab:real_data}).}
\label{tab:used_priors}
\end{table}
\begin{table}
\centering
\tabcolsep=0.11cm
\begin{tabular}{c|cccc|cccc|cccc}

\hline\hline
& \multicolumn{4}{c}{\textbf{HPV-6}} & \multicolumn{4}{c}{\textbf{HPV-11}} & \multicolumn{4}{c}{\textbf{combined HPV-6 and -11}}\\ \hline\hline
&  \multicolumn{2}{c}{\textbf{males}}& \multicolumn{2}{c}{\textbf{females}} &\multicolumn{2}{c}{\textbf{males}}& \multicolumn{2}{c}{\textbf{females}}&\multicolumn{2}{c}{\textbf{males}}& \multicolumn{2}{c}{\textbf{females}}\\
\hline\hline
\textbf{no.} & \textbf{\%} & \textbf{95\%CI} & \textbf{\%} & \textbf{95\%CI} &  \textbf{\%} & \textbf{95\%CI} & \textbf{\%} & \textbf{95\%CI}&  \textbf{\%} & \textbf{95\%CI} & \textbf{\%} & \textbf{95\%CI}\\
\hline
1  & 0.6  & 0.0-3.3   & 7.0   & 3.4-12.6   & 1.2  & 0.1-4.3   & 1.4 & 0.2-5 & 1.2 & 0.1-4.3  & 7.7 & 3.9-13.4 \\
2  & 9.6  & 5.9-14.4  & 12.2  & 7.5-16.9   & 7.2  & 4.1-11.6  & 4.9 & 2.1-7.7 & 13.4 & 9.1-18.8 & 18.2 &13.6-23.6\\
3  & 9.6  & 5.9-14.4  & 17.7  & 13.0-22.4  & 7.2  & 4.1-11.6  & 4.0 & 1.2-6.8 & 13.4 & 9.1-18.8  & 18.2 &13.6-23.6 \\
4  & 15.1 & 10.1-21.4 & 22.0  & 17.6-27.1  & 7.6  & 4.1-12.6  & 6.4 & 3.9-9.7 & 17.4 & 12.1-24.0 & 23.6 &19.0-28.7\\
5  & 15.1 & 10.1-21.4 & 22.0  & 17.6-27.1  & 7.6  & 4.1-12.6  & 6.4 & 3.9-9.7  & 17.4 & 12.1-24.0  & 23.6 &19.0-28.7\\
6  & 15.4 & 9.9-22.4  & 18.8  & 14.4-23.7  & 9.1  & 4.9-15.0  & 11.8& 8.3-16.1 & 20.3 & 14.0-27.8  & 24.0 &19.1-29.3\\
7 & 15.4 & 9.9-22.4  & 18.8  & 14.4-23.7  & 9.1  & 4.9-15.0  & 11.8& 8.3-16.1  & 20.3 & 14.0-27.8   & 24.0 &19.1-29.3\\
8 & 12.9 & 8.0-19.4  & 17.7  & 12.1-24.6  & 7.5  & 3.8-13.0  & 4.4 & 1.8-8.9 & 15.6 & 10.2-22.5 & 19.0 & 13.2-26.0\\
9  & 12.9 & 8.0-19.4  & 17.7  & 12.1-24.6  & 7.5  & 3.8-13.0  & 4.4 & 1.8-8.9 & 15.6 & 10.2-22.5   & 19.0 & 13.2-26.0\\
\hline\hline
\end{tabular}
\caption{Seroprevalence data for Australia as percentages of the whole male or female population (as reported in \cite{Newall2008}).}
\label{tab:seroprevalence_AU}
\end{table}

\begin{table}
\centering
\begin{tabular}{cc|cccc}
\hline\hline
\multicolumn{2}{c}{\textbf{age-group}}&  \multicolumn{2}{c}{\textbf{males}}& \multicolumn{2}{c}{\textbf{females}} \\ \hline \hline
\textbf{no.} & \textbf{ages, y.o.}  & \textbf{mean} & \textbf{95\%CI} & \textbf{mean} & \textbf{95\%CI} \\
\hline\hline
1 & 15-19 & 1.66  & 0.46-2.86 & 7.28   & 4.18-10.38    \\
2 & 20-24 & 6.27  & 3.77-8.77  & 8.61  & 5.61-11.61  \\
3 & 25-29 & 7.4  & 4.4-10.4  & 6.37  & 3.77-8.97  \\
4 & 30-34 & 4.64 & 2.44-6.84 & 4.33  & 2.13-6.53  \\
5 & 35-39 & 2.34 & 1.34-3.34 & 2.19  & 1.19-3.19   \\
6 & 40-44 & 2.34 & 1.34-3.34  & 2.19  & 1.19-3.19    \\
7 & 45-49 & 0.83 & 0.33-1.33 & 0.48  & 0.08-0.88  \\
8 & 50-54 & 0.83 & 0.33-1.33 & 0.48  & 0.08-0.88   \\
9 & 55-59 & 0.83 & 0.33-1.33 & 0.48  & 0.08-0.88 \\
\hline\hline
\end{tabular}
\caption{Genital warts incidence data for Australia measured as the mean number of new cases per 1000 persons per year (as reported in \cite{Pirotta2009}).}
\label{tab:GW_incidence_AU}
\end{table}

\begin{table}[h]
\centering
\begin{tabular}{cccccccccc}
\hline\hline
 &\multicolumn{9}{c}{age-group}\\
\hline\hline
no. & 1 & 2 & 3 & 4 & 5 & 6 & 7 & 8 & 9\\
\hline\hline
age, y.o. & 15-19 & 20-24 & 25-29 & 30-34 & 35-39 & 40-44 & 45-49 & 50-54 & 55-59\\
\hline\hline
rate, $r_a$ & 5.28 & 6.06 & 4.37 & 2.57 & 1.61 & 1.43 & 1.00 & 1.00 & 1.00\\
\hline\hline
\end{tabular}
\caption{Relative new sexual partner acquisition rates by age-group}
\label{tab:relative_rate_by_age_group}
\end{table}

\begin{table}[ht!]
\centering
\begin{tabular}{ccccc}
\hline\hline
sexual activity-group & 1 & 2 & 3 & 4\\
\hline\hline
population in each group, \% & 60 & 27 & 11 & 2\\
\hline\hline
rate, $r_s$ & 1.00 & 4.76 & 24.83 & 105.67\\
\hline\hline
\end{tabular}
\caption{Relative new sexual partner acquisition rates by sexual activity-group}
\label{tab:relative_rate_by_sex_group}
\end{table}

\end{document}